\documentclass{article}

\usepackage{caption}
\usepackage{subcaption}
\usepackage{arxiv}
\usepackage{graphicx}
\usepackage[utf8]{inputenc} 
\usepackage[T1]{fontenc}    
\usepackage{hyperref}       
\usepackage{url}            
\usepackage{booktabs}       
\usepackage{amsfonts}       
\usepackage{nicefrac}       
\usepackage{microtype}      
\usepackage{lipsum}
\usepackage[square, numbers, comma, sort&compress]{natbib}
\usepackage{url}

\title{Semantic Segmentation of Skin Lesions using a Small Data Set}

\author{
  Beril~Sirmacek, Max Kivits \\
  Robotics and Mechatronics\\
  University of Twente\\
  Enschede, The Netherlands \\
  \texttt{b.sirmacek@utwente.nl} \\
}

\begin{document}
\maketitle

\begin{abstract}
Early detection of melanoma is difficult for the human eye but a crucial step towards reducing its death rate. Computerized detection of these melanoma and other skin lesions is necessary. The central research question in this paper is \textit{"How to segment skin lesion images using a neural network with low available data?"}. This question is divided into three sub questions regarding best performing network structure, training data and training method. First theory associated with these questions is discussed. Literature states that U-net CNN structures have excellent performances on the segmentation task, more training data increases network performance and utilizing transfer learning enables networks to generalize to new data better.

To validate these findings in the literature two experiments are conducted. The first experiment trains a network on data sets of different size. The second experiment proposes twelve network structures and trains them on the same data set. The experimental results support the findings in the literature. The FCN16 and FCN32 networks perform best in the accuracy, intersection over union and mean BF1 Score metric. Concluding from these results the skin lesion segmentation network is a fully convolutional structure with a skip architecture and an encoder depth of either one or two. Weights of this network should be initialized using transfer learning from the pre trained VGG16 network. Training data should be cropped to reduce complexity and augmented during training to reduce the likelihood of overfitting.

\end{abstract}


\section{Introduction}

\label{ch:introduction}

A skin lesion is defined as damage or abnormality found in the skin, for instance chronic skin diseases like psoriasis or melanoma, a type of skin cancer. For melanoma early detection is crucial as it can be used to avoid most deaths caused by this type of cancer, and most skin cancers in general \cite{MelanomaFacts}\cite{Geller2015FocusDeaths}. When detected in an early stage a simple excision is a curative treatment and the 5-year relative survival rate is 97\%. This drops to less than 50\% for stage III melanoma, decreasing exponentially with each subsequent stage \cite{SurvivalStage}. 

Visually early stage melanomas are difficult to distinguish from their benign counterparts. This leads to many missed or false biopsies, causing great personal and financial harm. The importance of early and accurate melanoma detection is obvious. Experts with specialized equipment can very accurately diagnose even very early stage melanomas. However, this equipment is very expensive and not enough experts are available. To solve this problem digital skin imaging techniques have been developed, increasing the efficiency and accuracy of diagnosis and reducing the need for expert personnel.

In this paper such a digital skin imaging technique is presented. A neural network is designed to semantically segment images of skin lesions, labeling each pixel as either 'Lesion' or 'Skin'. A data set has been generated at the University of Twente (UT). This set contains 41 images taken by a regular smartphone camera. These images display different body parts containing skin lesion area. Machine learning networks require large amounts of data to train. State of the art networks train for weeks on ImageNet, a database containing over 14 million images \cite{HeDeepRecognition} \cite{Russakovsky2015ImageNetChallenge}. Compared to ImageNet the size of the data set provided by the University is exceptionally small.

Machine learning is a rapidly growing discipline. The number of fields machine learning can be successfully applied in is only limited by the amount of available data in those fields. In an attempt to find solutions to this problem and construct a functioning network for the UT, maximizing performance with a small data set is the central theme of this paper. The research question is: \textit{"How to segment skin lesion images using a neural network with low available data?"}

In order to better understand the problem and focus research multiple sub questions are formulated:
\begin{enumerate}
\label{int:sub}

\item How does training data impact specific and general segmentation performance?

\item How does network structure impact specific and general segmentation performance? 

\item How does training method impact specific and general segmentation performance?

\end{enumerate}


In order to give insight into these sub questions two experiments are conduced.
First a network is trained on different data sets. To find the best performing network their segmentation performances are compared. The second experiment compares the segmentation performance of different network structures trained on the same data set. To drawn conclusions their performances are compared. The second experiment also yields observations regarding the third sub question. Some of the network stuctures trained in this experiment have weights partially transfered from the pre trained VGG16 network. Conclusions on the impact of training methods are drawn from the comparison of these networks to randomly initialized networks. 

The second section gives background information regarding segmentation and neural network theory, discussing the optimal structure for the segmentation task. Overfitting and strategies to prevent it from occuring are discussed. The difference between randomly initializing network weights and transfering them from pre-trained networks is explained. Strategies to deal with a sparse data set are given. In the final part of this section the different metrics used to compare network performance are explained.

The third section discusses the experimental setup of the conducted experiments. The network structures used are given. Information of the metrics by which performance will be measured is also provided. Two publicly available data sets are used in this paper. All used data sets are listed below and specified in more detail in the third section \ref{ch:method}.

\begin{itemize}
    \item 41 Skin lesion images provided by the University of Twente
    \item 126 Skin lesion images made publicly available by E. Nasr et al. \cite{Nasr-EsfahaniDenseSegmentation}
    \item 10 Skin lesion images publicly available in the ISIC Archive \cite{ISDISISICArchive}. 
\end{itemize}

Section four lists all the results obtained in the experiments done in this paper. Discussion of these results can be found in section five. In this section, observations on the experimental results are made and conclusions regarding the research sub questions are drawn. 

Section six summarizes the observations made in section five and concludes the paper by answering the main research question.

\section{Background}
\label{ch:theory}

Semantics are the relation between words and symbols to their respective meanings. In computer vision, semantic segmentation is the practice of labeling each pixel of an image in order to obtain regions within that image that share the same high level features. For instance mapping each pixel of a driving scene image to either road, cars or pedestrian. Image segmentation is a fundamental task in computer vision because it enables machines to extract contextual information and make decisions based on it.

Traditionally semantic segmentation is done by means of pre processing methods like thresholding, clustering or edge detection \cite{Celebi2009LesionImages.} \cite{Sumithra2015SegmentationDiagnosis}. An extremely valued property of segmentation algorithms is their robustness to input variability. The aforementioned "handcrafted" methods usually struggle greatly in this area and require manual attention when input variations occur. 

Contemporary segmentation methods instead leverage deep learning techniques which are more resilient and effective. Steady increase in processing power and availability of labeled data have caused an increase in interest in the machine learning field.
Deep learning models based on neural networks have outperformed traditional methods in the annual ImageNet Large Scale Visual Recognition Challenge (ILSVRC) since 2012 \cite{Russakovsky2015ImageNetChallenge}. Today machine learning is used extensively in many fields such as computer vision, marketing and drug discovery \cite{Ching2018OpportunitiesMedicine}.

\subsection{Neural Network Layers}
\label{sc:networkLayers}
Artificial Neural networks (ANNs or simply "networks") are composed of connected layers of computational nodes called neurons that work together to optimize the network output. The first network layer is called the input layer. It takes the network input, which are usually a vector or matrix. The following layers are called hidden layers and the final layer is the output layer. 

 Connections between neurons are weighted. Each layer performs an operation on the output of the previous layer. A network input changes the neuron activations in the input layer. Neuron activations then propagate through the network layers until the output layer is reached. An optimization algorithm compares the network output with the desired output and determines the optimal weights for the neuron connections. This optimization process is called network training and is responsible for an ANNs' ability to 'learn'.

\begin{figure}[h!t]
	\begin{center}
		\includegraphics[width=\textwidth]{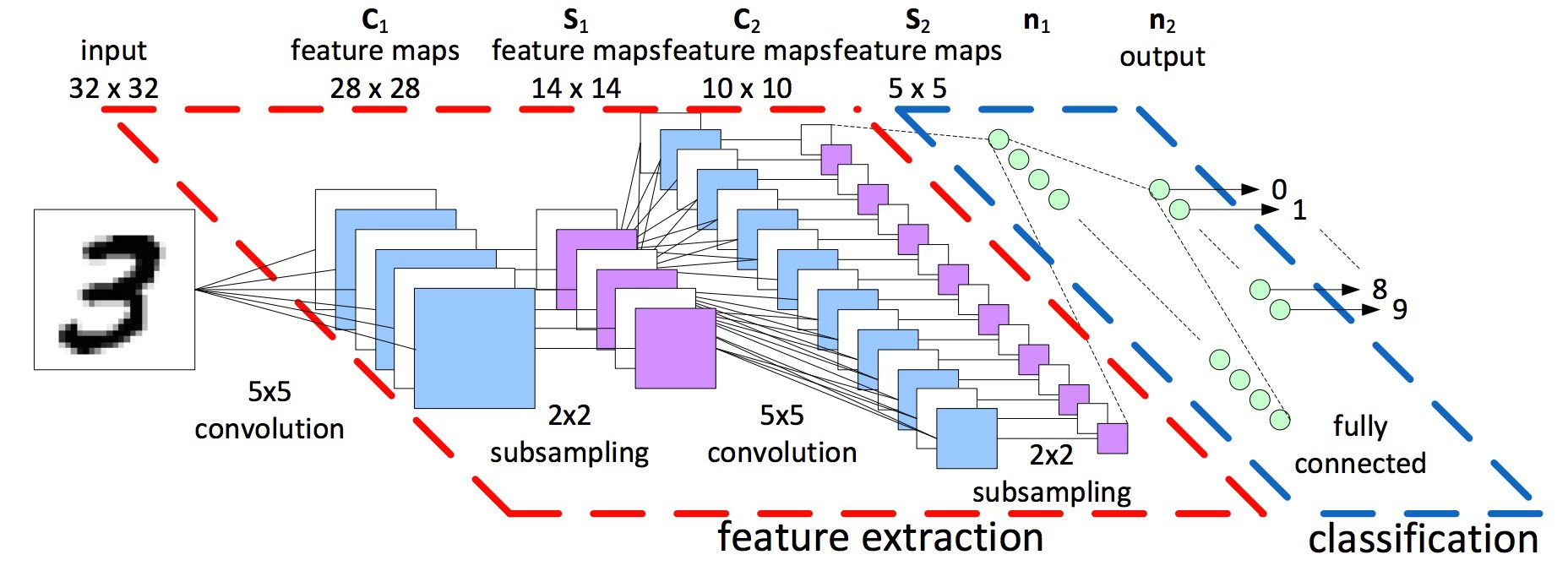}
	\end{center}
	\caption{Commonly used classification network structure, taken from \cite{PeemenLNCSPlatform}}
	\label{fig:basicClassNet}
\end{figure}

A lot of different neural network structures exist. A common type of computer vision network is the classification network. This network tries to classify an input image by calculating the probabilities of all the classes it is trained on. A general classification network is given in figure \ref{fig:basicClassNet}, this particular network is designed to classify digits from the MNIST data set. Most state of the art performances in computer vision challenges are achieved using convolutional neural networks(CNNs). These are a special type of network that use convolutional layers. Although naming convention suggests otherwise, these layers use 2D cross-correlation operations to extract image features from their input. This operation is defined as:

\[ Y_{i,j} = \sum _{\delta_{i},\delta_{j},k}^{} X_{i+\delta_{i},j+\delta_{j},k} \cdot K_{\delta_{i},\delta_{j},k} \]

\begin{figure}[h!t]
	\begin{center}
		\includegraphics[width=325pt]{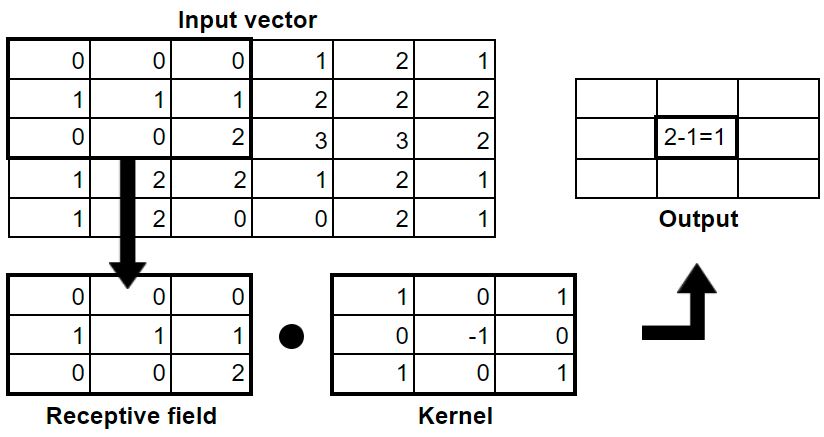}
	\end{center}
	\caption{Visualization of 3x3 Cross-correlation Operation }
	\label{fig:basicConv}
\end{figure}

For output $ Y_{i,j} $ at position $i,j$, with input $ X_{i,j}$, filter kernel $K$, input dimension $k$ and x,y filter length $\delta{i}$ and $\delta{j}$ respectively. A filter or kernel is moved over the entire input vector, producing a receptive field. The output of the layer is a dot product of the receptive field and the kernel. A visual representation of this operation is given in figure \ref{fig:basicConv}. The step size of this operation is defined as its stride. Before network training kernel size and stride are defined. Generally the kernel is randomly initialized.

\begin{figure}[h!t]
	\begin{center}
		\includegraphics[width=275pt]{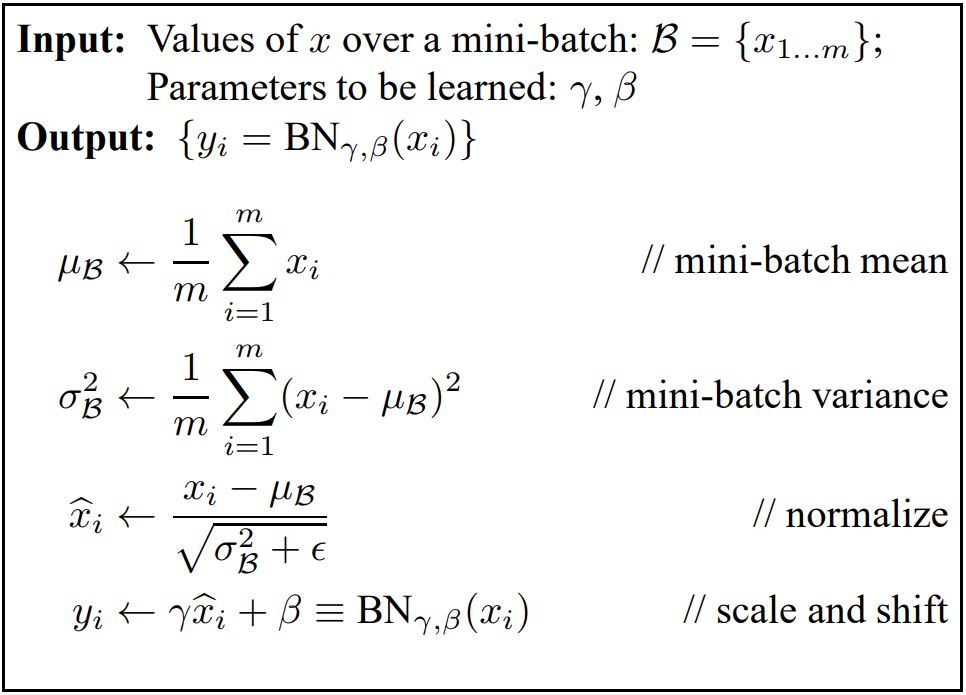}
	\end{center}
	\caption{Batch Normalizing Transform, applied to
activation $x$ over a mini-batch, taken from \cite{Ioffe2015BatchShift}}
	\label{fig:batchNorm}
\end{figure}

Another commonly used layer in CNNs is the batch normalization layer. This layer normalizes the output of previous layers by subtracting their batch mean and dividing by their variance, as seen in figure \ref{fig:batchNorm}. This is useful to reduce the influence of extremely high neuron activation values on the rest of the network. This enables networks to use a higher learning rate without the risk of overfitting and allows networks to reach a more optimal trained state. Only normalizing a layer can change the information it is representing, to compensate for this the total transform function of this layer has to be able to represent the identity transform \cite{Ioffe2015BatchShift}. In order to do this a learnable shift parameter $\beta$ and a scale parameter $\gamma$ are added to the batch normalization transform. 

\begin{figure}[h]
    \centering
    \begin{subfigure}[b]{200pt}
        \includegraphics[width=180pt]{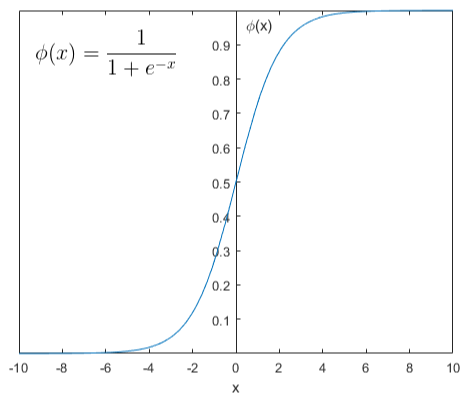}
        \caption{Sigmoid activation function}
        \label{fig:Sigmoid}
    \end{subfigure}
    \begin{subfigure}[b]{200pt}
        \includegraphics[width=180pt]{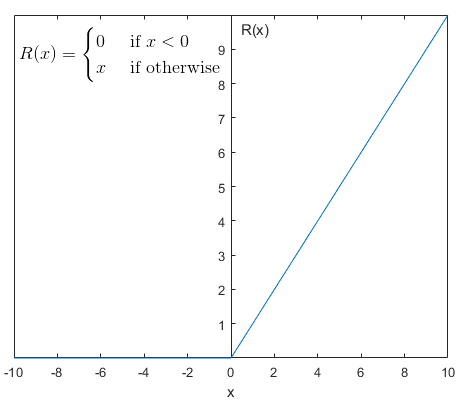}
        \caption{ReLu activation function}
        \label{fig:ReLu}
    \end{subfigure}
    \caption{Activation Functions}\label{fig:ActivationFunc}
\end{figure}

Activation layers are usually implemented after normalization layers. These layers transform their input by applying an activation function to them. Inspired by biological neurons, activation functions aim to mimic the threshold potential during neural excitement. These functions are usually non-linear. An example is the Sigmoid activation function, figure \ref{fig:Sigmoid}. This function bounds the input to either 0 or 1 at very small or high activation levels respectively. This function was heavily used but recently the rectifier linear unit or ReLu, figure \ref{fig:ReLu}, is gaining popularity. Compared to the Sigmoid, ReLu is computationally more efficient. It also solves the Sigmoids' vanishing gradient problem. This is a problem that occurs during training of the network where the Sigmoid function causes the optimization algorithm to stall. 

\begin{figure}[h!t]
	\begin{center}
		\includegraphics[width=275pt]{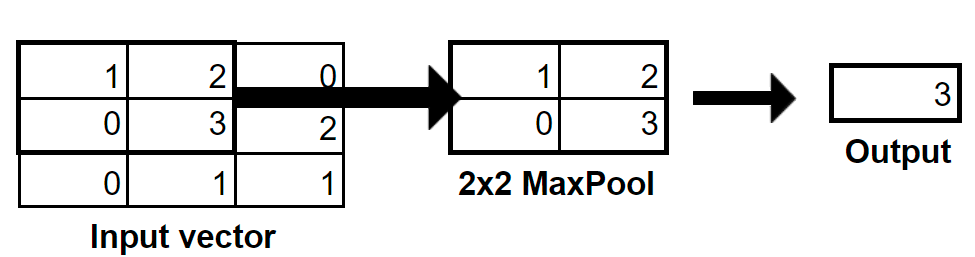}
	\end{center}
	\caption{Visualization of a 2x2 Max Pooling operation }
	\label{fig:MaxPool}
\end{figure}

\begin{figure}[h!t]
	\begin{center}
		\includegraphics[width=300pt]{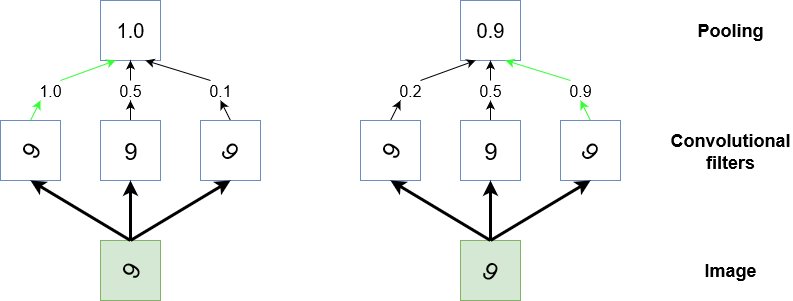}
	\end{center}
	\caption{Visualization of Generalization when combining Convolutional and Max Pooling layers \cite{AndyThomas2017ConvolutionalLearning}}
	\label{fig:PoolDemo}
\end{figure}

Another extremely common layer in CNNs is the pooling layer. This layer reduces spatial dimensions by downsampling or 'pooling' its input and producing a single output from it. The Max Pooling layer is the most common, illustrated in figure \ref{fig:MaxPool}. Pooling servers two purposes. It reduces computational difficulty by reducing the number of parameters in the network. Pooling also aids in object detection by generalizing high resolution data to lower resolution information \cite{AndyThomas2017ConvolutionalLearning}, illustrated in figure \ref{fig:PoolDemo}. 

\subsection{Overfitting \& Underfitting}
\label{sc:overfitting}
\begin{figure}[h!t]
	\begin{center}
		\includegraphics[width=420pt]{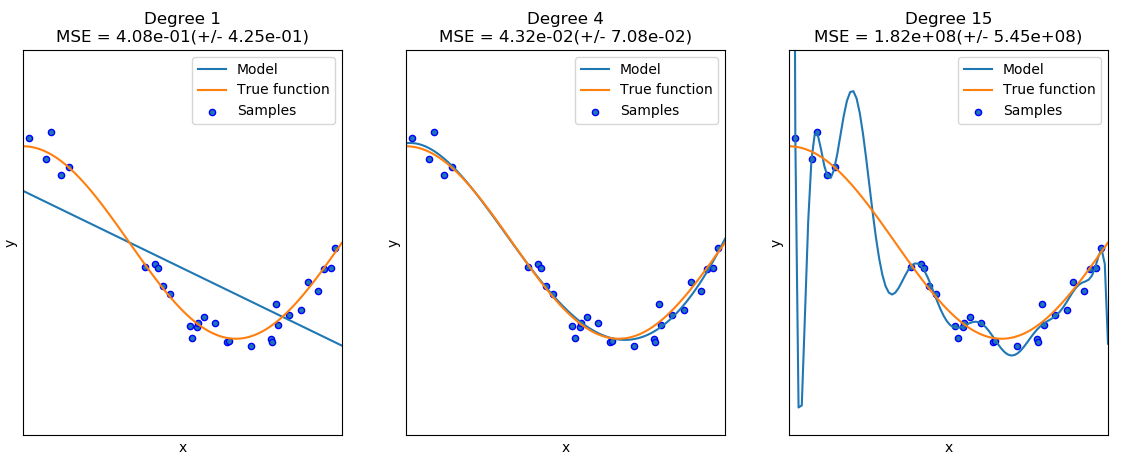}
	\end{center}
	\caption{Under- and Overfitting a polynomial curve}
	\label{fig:overfit}
\end{figure}

Overfitting is defined as: "The production of an analysis which corresponds too closely or exactly to a particular set of data, and may therefore fail to fit additional data or predict future observations reliably", from the Oxford dictionary \cite{OverfittingDictionaries}. Underfitting occurs when a statistical model assumes a  relationship between data points that is too simple. Figure \ref{fig:overfit} shows a visualization of over- and underfitting a polynomial on a small data set. MSE is the mean squared error on the validation set \cite{UnderfittingOverfitting}.

Overfitting in neural networks manifests as poor performance on the test set compared to the training data set. This signifies a loss of generalization. Using a small amount of training data significantly increases likelihood of overfitting. Increasing network width has been shown to reduce the effects of overfitting \cite{Zagoruyko2016WideNetworks}. The generalization effect caused by the combination of convolution and pooling layers discussed in section \ref{sc:networkLayers} also combats overfitting.

Regularization or \textit{Weight Decay} is an effective method to reduce overfitting \cite{Schmidhuber2014DeepOverview}. Many different regularization techniques exist. Most work by adding a penalty term to the loss function of a network. Two common and similar techniques are L1 and L2 regularization. L2 or \textit{Ridge Regression} adds a squared magnitude term of the weights vector. L1 or \textit{Lasso Regression} adds only an absolute value term of the weights vector. Compared to L1, loss functions with a L2 term are more influential to statistical outliers.In practice L1 and L2 are commonly used when training on large amounts of data. 

Early stopping is an often implemented regularization technique. Up to a point the training process improves the networks performance on data outside the training and test sets. Past this point any improvements made on the training and testing set comes at the expense of network generalization performance. The aim of early stopping is to stop network training at this turning point by providing certain conditions. An often used early stopping condition is requiring improvement in validation accuracy within the last X validation performance tests. 

Dropout is another effective regularization technique that aims at reducing overfitting by preventing complex co-adaptations on training data. Networks with dropout ignore a ratio of neurons activations during training proportionate to its dropout rate. Batch normalization is often favored over dropout in convolutional architectures. \cite{Ioffe2015BatchShift}.

\subsection{Network Structure}

\begin{figure}[h]
	\begin{center}
		\includegraphics[width=300pt]{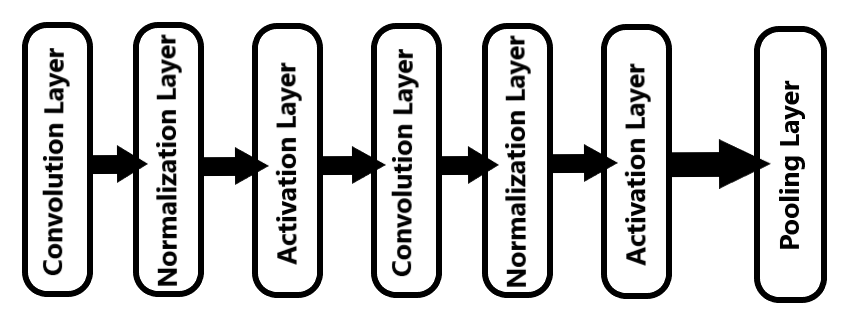}
	\end{center}
	\caption{Example of multiple layer segmentation network structure section}
	\label{fig:section}
\end{figure}

\begin{figure}[h]
	\begin{center}
		\includegraphics[width=300pt]{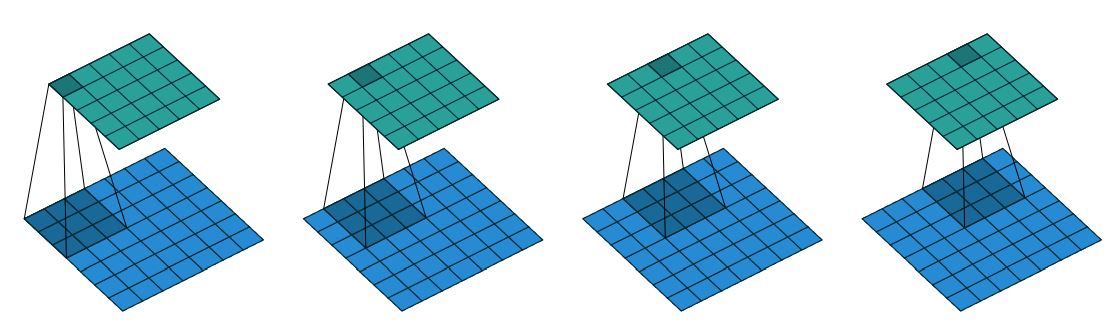}
	\end{center}
	\caption{Visualization of transpose convolution operation, taken from \cite{Dumoulin2016ALearning}}
	\label{fig:transConv}
\end{figure}

\begin{figure}[h]
	\begin{center}
		\includegraphics[width=300pt]{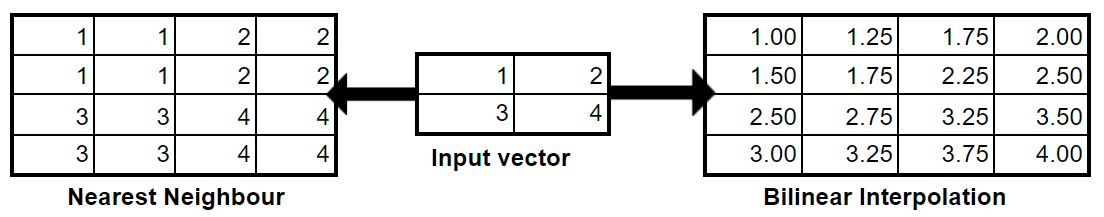}
	\end{center}
	\caption{Visualization of upsampling algorithms}
	\label{fig:upsampling}
\end{figure}

\begin{figure}[h]
	\begin{center}
		\includegraphics[width=300pt]{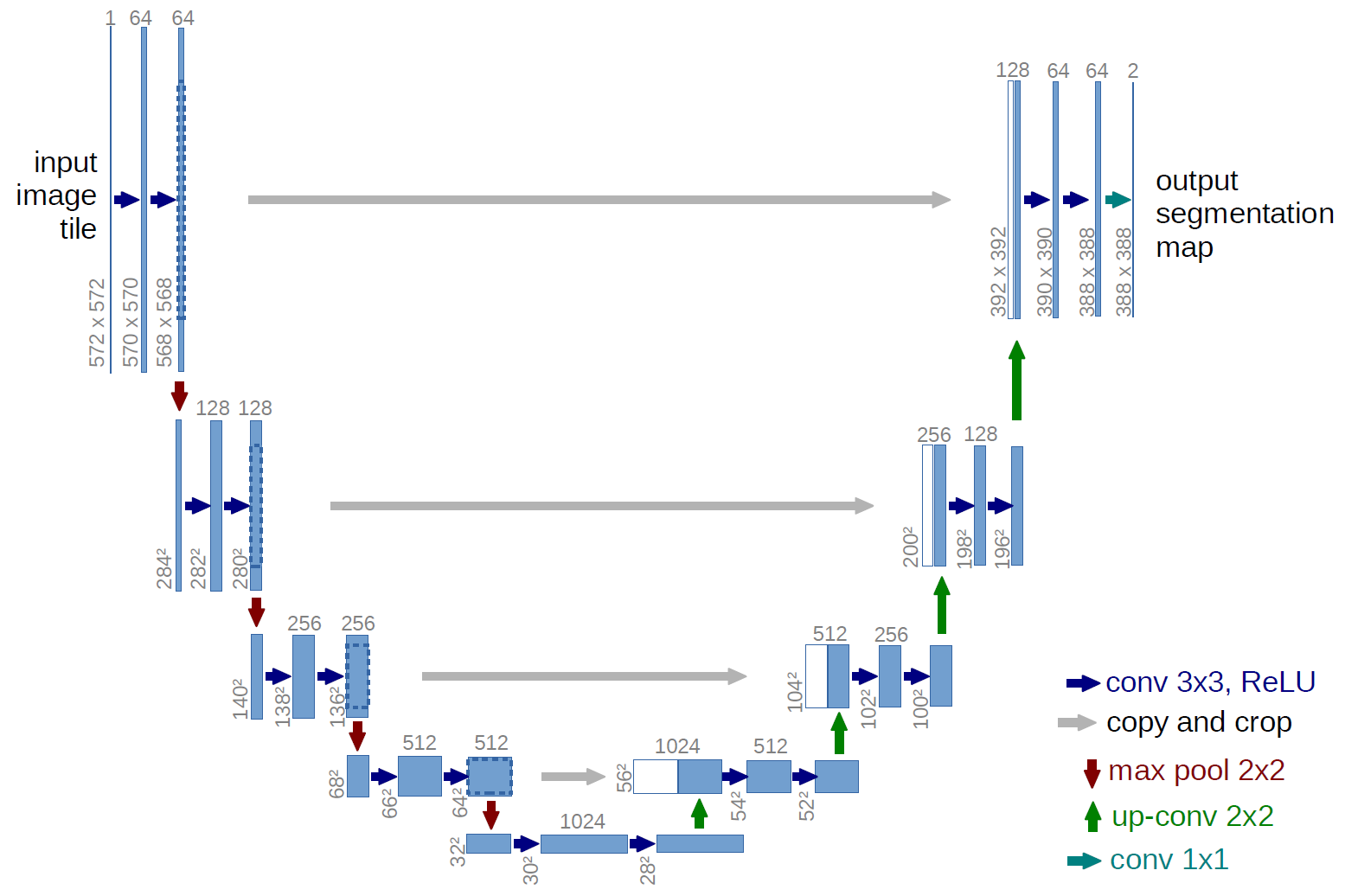}
	\end{center}
	\caption{High level U-net segmentation network structure, taken from \cite{Ronneberger2015U-Net:Segmentation}}
	\label{fig:segNetStruct}
\end{figure}

A network structure hat performs very well on segmentation tasks is the deconvolutional or U-net structure \cite{Ronneberger2015U-Net:Segmentation}. This structure is build up from a combination of repetitive sections. One section usually contains a combination of convolution, batch normalization and activation layers followed by a pooling layer. An example section is given in figure \ref{fig:section}.
U-net structures combine these sections to form an encoding and decoding structure, as illustrated in figure \ref{fig:segNetStruct}. Encoding is the descending part of the structure. This aims at extracting low level information like textures by reducing resolution using convolution and pooling operations. Decoding is the ascending part of the structure. It acts as the transpose of an encoding layer, increasing the resolution until reaching the input image size. In the decoding structure all convolution layers are replaced by deconvolution layers that perform a transpose convolution operation illustrated in fig \ref{fig:transConv}. All Pooling layers are replaced by layers performing upsampling operations. Nearest neighbour and bilinear interpolation are the two most common upsampling algorithms, illustrated in figure  \ref{fig:upsampling}. 
\newline
U-net structures connect these encoding and decoding parts by means of a 'skip architecture', visible as grey arrows in figure \ref{fig:segNetStruct}. The intuitive reasoning behind implementing these connections is to combine semantic information from deep layer feature maps with environmental information from more shallow layers. This forces the network to look at environmental information in hopes of creating a network that is able to process contextual information. The total decrease of resolution within a network is defined as network depth. Every network layer also has a width, defined as the total amount of neurons within that layer. 
 
Although U-net structures generally perform very well on segmentation tasks it gives no further specific details like network depth, convolution size or network structure section composition. These specifications are usually called network parameters or \textit{Hyper-parameters}. Optimal network parameters depend on the task they are required to perform. Determining the optimal network depth is subject of much debate, with opinion split between favoring either network depth or width. Increase of either has been proven to yield increased network performance \cite{Simonyan2015VERYRECOGNITION} \cite{Zagoruyko2016WideNetworks}. Choosing network depth and width is always a consideration between performance and computational efficiency. An optimum between these two attributes is generally found by experimentation. 

Image input resolution is another parameters to consider. Higher input resolutions allow the network to detect features more accurately. This can cause a very significant increase in network performance. A major drawback of increasing input resolution is the major increase in computational cost. An increase input resolution causes a proportional increase in neurons in the input and subsequent layers. This greatly increases the number of parameters of the network and the computational cost of the training process.
Another parameter that is not directly defined by the U-net structure is the kernel size in convolution layers. Kernel size determines the resolution at which a convolution layer searches for features. Kernel size can range from 1x1 to the entire input resolution. 1x1 kernels look for features in singular pixels, disregarding any information contained in surrounding pixels. Kernels spanning the entire input range take all pixels into account when looking for features, likely skipping important details. In recent years researchers have started to favor 3x3 kernels over bigger kernels \cite{KrizhevskyImageNetNetworks} \cite{Simonyan2015VERYRECOGNITION} for image processing tasks.

In practice the optimal kernel size is found by experimentation. This extends to many Hyper-parameters. To my knowledge there exists no generic algorithm to \textit{a priori} determine the optimal network structure for a given problem. Change in a network parameter heavily influences other network parameters. 
There are however a few general parameters optimization techniques. A commonly used strategy is starting with the simplest possible network and using experimental validation to determine the optimal level of network complexity. Experiments are often done with the dropout parameter discussed in section \ref{sc:overfitting}. Experiments on network depth and width are also common practice. The metric used in these experiments is the ratio of network performance to computational cost, with more optimal networks scoring higher.

\subsection{Training Methods}
The training process is responsible for a networks' performance. Network training aims to minimize a loss function associated with the network. This loss function depends on neuron weights and other learnable parameters. This function is generally composed of an error and a regularization term. The error term evaluates the performance of a network by comparing the output to the training data set. The regularization term prevents overfitting by forcing a network to favor less complex solutions over more complex ones. 

\begin{figure}[h!t]
    \centering
    \begin{subfigure}[b]{200pt}
        \includegraphics[width=180pt]{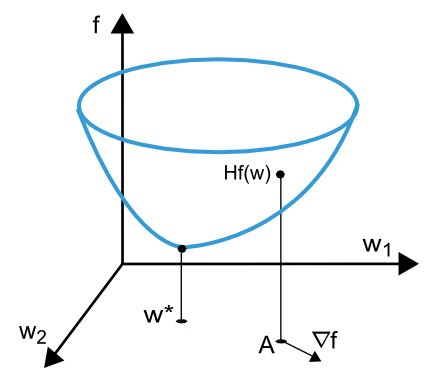}
        \caption{Simple loss function, \cite{AlbertoQuesada5Network}}
        \label{fig:lossSimple}
    \end{subfigure}
    \begin{subfigure}[b]{200pt}
        \includegraphics[width=180pt]{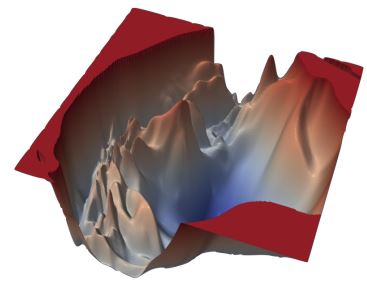}
        \caption{ResNet-110 without skip connections, \cite{LiVisualizingNets}}
        \label{fig:lossHard}
    \end{subfigure}
    \caption{Loss functions}\label{fig:ActivationFunc}
\end{figure}

Figure \ref{fig:lossSimple} contains a hypothetical loss function $f(w)$. This loss function only depends on two weights: $W_{1}$ and $W_{2}$. The training process aims at reaching the global minimum of the loss function, in this case the point $W^{*}$. At any point A the first and second derivatives of the function can be calculated. They can respectively be written as the gradient vector $\nabla _{i}\textup{f(w)}$ and Hessian matrix $H_{i,j}\textup{f(w)}$:

\begin{equation}
    \nabla _{i}f(w)= df/dw_{i}\qquad (i = 1,...,n) 
\end{equation}
\begin{equation}
    H_{i,j}f(w) = d^2f/dw_{i} \cdot dw_{j} \qquad (i,j = 1,...,n)
\end{equation}

Most practical networks have a lot more learnable parameters and contain many non-linearities, complicating the loss function. Figure \ref{fig:lossHard} serves to illustrate this, displaying the loss landscape of the very deep ResNet-110 network \cite{HeDeepRecognition}. Due to the complex nature of loss functions and the non-linearities within them it is not possible to find closed training algorithms to obtain the minima. Instead weights are often initialized randomly and the network is trained using iterative methods. Finding the global minima is still extremely difficult because of the many local minima present in more complex loss functions. Often training stops when a relatively small local minima is reached.

Gradient or stochastic gradient descent (SGD) is the most commonly used iterative algorithm. Each iteration is formulated as equation \ref{eq:gradientDescend}, starting at point $W_{0}$. Each iteration the gradient vector is calculated and a step is taken in the reverse direction of the gradient $g$. Since the gradient vector gives information of the direction of quickest ascend of a function, moving in the reverse direction would yield the quickest descend. The size of the step taken every iteration is based on the training rate $\eta$ which usually decreases in size as training progresses.  Gradient descend is a first order algorithm as it only requires information of the gradient vector, reducing computational cost. This difference causes gradient descend to be favored when training large networks. 

A common addition to gradient descent is momentum. SGD with momentum updates the weights of a network it is optimizing based on a moving average of past iterations. Instead of updating weights every iteration. This addition almost always increases performance.  

\begin{equation}
\label{eq:gradientDescend}
    W_{i+1} = W_{i} - g_{i} \cdot \eta_{i},\qquad i=0,1,...
\end{equation}

Another training algorithm is the Newton's method. This is a numerical analysis method can be used to find 0 crossings of functions. In network training it functions as a second order algorithm, requiring information of both the gradient vector and the hessian matrix. Each iteration follows equation \ref{eq:newton}, with $H_{i}$ the Hessian of the loss function $f(w_{i})$.  The main advantage of second order methods is a significant increase in convergence speed. A disadvantage of these methods is their information requirement of the Hessian, increasing computational cost. There exits other algorithms like the Conjugate gradient and Quasi-Newton method that aim at reducing this computational cost whilst still making accurate steps. 

\begin{equation}
\label{eq:newton}
    W_{i+1} = W_{i} - (H_{i}^{-1} \cdot g_{i}) \cdot \eta_{i}, \qquad i=0,1,...
\end{equation}

In CNNs that process images, the first few convolution layers usually develop similar kernels during training. These kernels are called Gabor filters and appear to be similar to the filter response of human cortical cells \cite{Marcelja1980MathematicalCells}. This similarity between image processing networks suggests the feasibility of a training method called transfer learning. Evidence shows that copying weights from a pre trained network,  instead of randomly initializing them, is beneficial to network performance. Even when the pre trained network is designed for very different input data and only a few layers are transferred \cite{YosinskiHowNetworks}.

\subsection{Data manipulation}
Training a network on a bigger data set is an effective way to reduce overfitting and increase its generalization performance. Whether there is an upper limit to this effect or not is a topic of discussion. Papers written by A. Halvey and M. Banko et al. claim data set size is the main limiting factor of statistical model effectiveness \cite{2009TheData} \cite{BankoScalingDisambiguation}. Z. Xiangxin et al. have a contrasting opinion and conjecture that "the greatest gains in detection performance will continue to derive from improved representations and learning algorithms that can make efficient use of large datasets."\cite{ZhuDoData}.

In any case it remains beneficial to acquire more training data.  Qualitative and labeled data is seldom publicly available. Another method of acquiring more training data is using data augmentation. Existing training data can be scaled, rotated, translated or otherwise transformed to produce new data. Training a network on augmented training data increases its generalization performance by forcing it to learn from contextual information. 

Another data augmentation technique is adding different kinds of noise to training data. This almost always increases network performance and sometimes even causes a network to converge faster\cite{LuoDeepNoise}.

A third strategy is to generate more data using \textit{Generative Adversarial Networks} (GANs). GANs are a machine learning model used to generate images from noise. GANs can be trained to generate images similar to the data they are trained on. Transfer learning can be applied to a pre trained GAN, configuring it to generate useful training images. This is a very effective method of increasing training data set size and increase network performance, according to L. Sixt et al.\cite{SixtRENDERGAN:DATA}.   
\subsection{Pre processing}
Using filters to enhance network training data is another effective data augmentation technique. The goal of this technique is to increase visibility of textures and other features, reducing the difficulty of training a network to identify those features.  

\subsection{Performance Metrics}
\begin{figure}[h!t]
	\begin{center}
		\includegraphics[width=200pt]{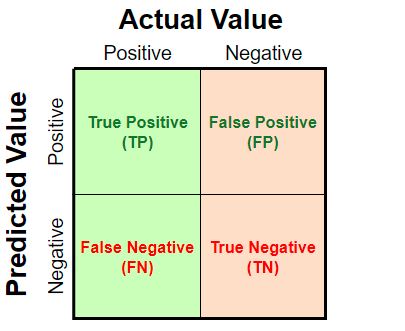}
	\end{center}
	\caption{A Confusion Matrix}
	\label{fig:TPFN}
\end{figure}

Segmentation performance of networks can be measured using different metrics. One of these metrics is the confusion matrix, displayed in figure \ref{fig:TPFN}. Another useful metric is label, mean and weighted mean accuracy. Equation \ref{eq:labacc} shows how label accuracy is calculated, with P total number of actual positives. Mean accuracy is the average accuracy of all labels. Weighted accuracy is the weighted accuracy of all labels. 

\begin{equation}
    \textup{ACC} = \frac{\textup{TP}}{\textup{P}}
    \label{eq:labacc}
\end{equation}

\begin{figure}[h!t]
	\begin{center}
		\includegraphics[width=200pt]{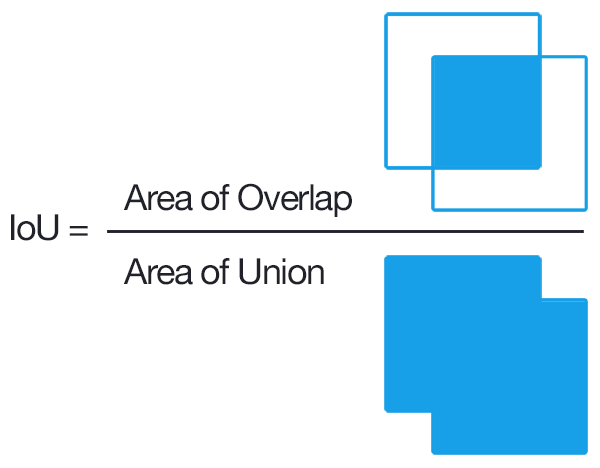}
	\end{center}
	\caption{Visualization of the Intersection over Union metric, from \cite{IntersectionPyImageSearch}}
	\label{fig:IoU}
\end{figure}

Intersection over Union (IoU) is a commonly used metric. It is the ration of the intersection of the actual and predicted label to their union. IoU is visualized in figure \ref{fig:IoU}. When looking at a network with a large accuracy its results can still look quite far from optimal. This deceiving effect of the accuracy metric is why IoU is often used when comparing segmentation networks.

A third metric is the F1 score. It uses the notion of precision and recall. Precision is a measure of the number of true positives to the number of total perceived positives. It is a measure of the amount of false positives a system generates. The precision metric increases in importance when the consequences of false positives are very negative. Recall is the number of true positives divided by the number of true positives and false negatives. A low recall indicates a large number of false negatives.

Precision (PPV) and recall (TPR) are calculated as:
\begin{equation}
\textup{PPV} = \frac{\textup{TP}}{\textup{TP+FP}}
    \label{PPV}
\end{equation}
\begin{equation}
\textup{TPR} = \frac{\textup{TP}}{\textup{TP+FN}}
    \label{TPR}
\end{equation}

F1 score is the harmonic mean of precision and recall. This gives information of the balance a system strikes between both metrics. 
\begin{equation}
\textup{F1} = 2 \cdot \frac{\textup{PPV} \cdot \textup{TPR}}{\textup{PPV+TPR}}
    \label{PPV}
\end{equation}

\section{Methods}

\label{ch:method}
\begin{figure}[h!t]
	\begin{center}
		\includegraphics[width=420pt]{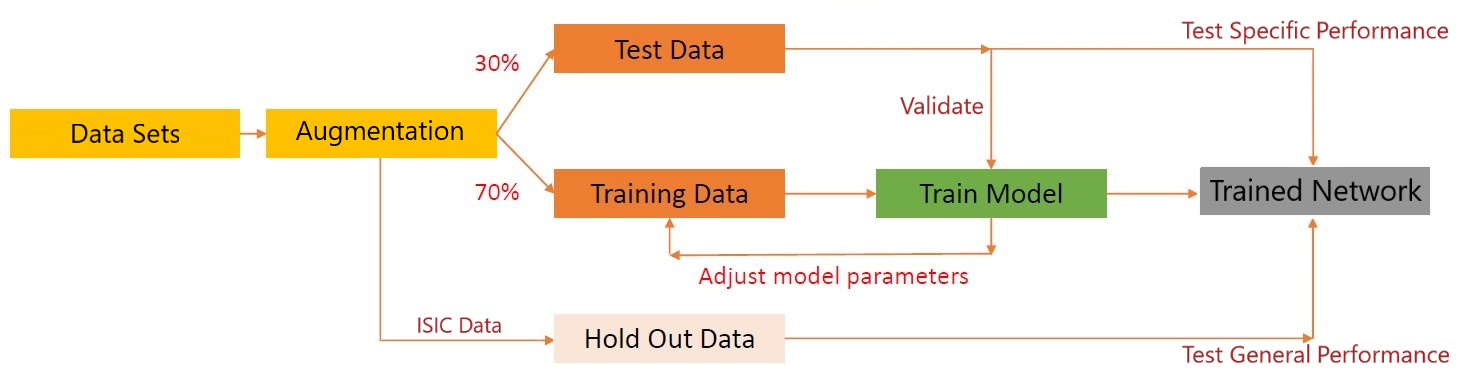}
	\end{center}
	\caption{Block Diagram of Experimental Setup}
	\label{fig:method}
\end{figure}

In order to answer the research question and sub-questions Two experiments are conducted. The Training Data experiment serves to examine the impact of training data on specific and general segmentation performance. The impact of network structure and training method is examined in the Network Structure experiment. Figure \ref{fig:method} shows a block diagram of the experimental setup used in both experiments.

All preprocessing, network training, analysis and post-processing are done using the Matlab R2018b software \cite{MathWorksMathWorksMATLAB}. All experiments are done on a desktop computer, specifications are listed in table \ref{tab:specs}

\begin{table}
\centering
\setlength{\aboverulesep}{0pt}
\setlength{\belowrulesep}{0pt}
\begin{tabular}{ll} 
\toprule
\textbf{Component} & \textbf{Specification}  \\ 
\hline
CPU                                               & i5~ 6600k @3.5GHz       \\
RAM             & 16GB @1200MHz           \\
GPU                                               & GTX 970 4GB Vram        \\
OS              & Windows 10 pro 64-bit   \\
\bottomrule
\end{tabular}
\caption{Specifications of the hardware used in this experiment}
\label{tab:specs}
\end{table}
\subsection{Data Sets}

\label{sc:dataset}
The definition of a skin lesion is very broad and no additional information is given regarding the type of lesions the network should be optimized for. Because of this, combined with the scarcity of properly diagnosed skin lesion data sets, the network is trained to detect general skin lesions. This allows the use of publicly available skin lesion data sets, increasing the amount of training data available.

The data used in these experiments is gathered from three different sources. All images are RGB. he DFCN data set is made publicly available by the E. Nasr et al. \cite{Nasr-EsfahaniDenseSegmentation}. The ISIC data set is taken from the ISIC Archive under public CC0 license. The UT image set is built at the University of Twente. Labels for this image set are made using the MatLab \href{https://nl.mathworks.com/help/vision/ref/imagelabeler-app.html}{Image Labeler}. 
Data set information can be found below in table \ref{tab:dataset}.

\begin{table}
\centering
\begin{tabular}{llllll} 
\toprule
\textbf{Data set} & \textbf{Author}  & \textbf{Used Images} & \textbf{Image Size} & \textbf{Reference} & \textbf{Link}  \\ 
\hline
UT                                               & Lab              & 41                        & 2560 x 1920     &     $\times$              &     $\times$         \\
 DFCN           & E. Nasr et al. & 126                       & 600 x 400      &  \cite{Nasr-EsfahaniDenseSegmentation}     &  \href{https://github.com/ebrahimnasr/DFCN/}{DFCN Github}              \\
ISIC                                             & ISDIS \& IDS            & 10                     & [1022-2592] x [767-1944]     &  \cite{ISDISISICArchive}                   &  \href{https://www.isic-archive.com}{ISIC Archive}              \\
\bottomrule
\end{tabular}
\caption{Data set Specifications}
\label{tab:dataset}
\end{table}

The Dermnet, DermnetNZ, Dermquest and the Brazilian Dermatology Atlas Databases were contacted without reply. The ISIC database \cite{ISDISISICArchive} contains vast amount of images. However due to time and computational power constraints only a small subset is used.

\begin{figure}[h]
    \centering
    \includegraphics[width=420pt]{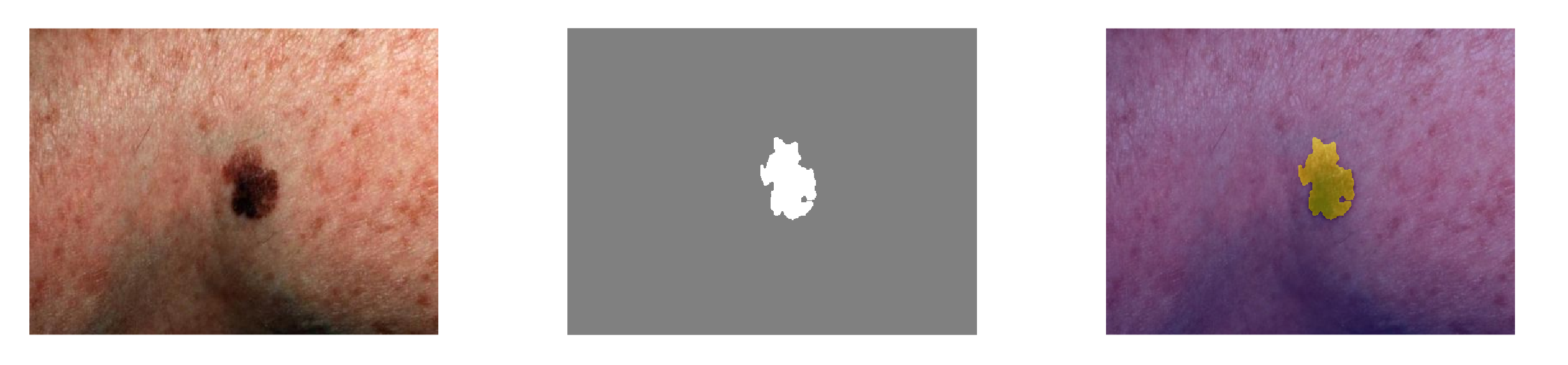}
    \caption{Overlay of a Label and corresponding Image}
    \label{fig:imglab}
\end{figure}

In order to reduce computational cost all data is resized to 360 x 480 prior to training. Labels are grayscale files. Each label pixel is assigned a labelPixelID of 0,1 or 2 representing "Background", "Skin" or "Lesion" respectively. Compared to the other dermatological data sets, our smartphone images from the UT data set contain significantly more background. Training the network to be able to separate background from useful input would significantly increase training difficulty. To combat this during network training all image pixels that are classed as "Background" in the corresponding label are disregarded. An example image label pair and their overlay is given in figure \ref{fig:imglab}.

During network training images are augmented in order to reduce overfitting. Augmentation details can be found in table \ref{tab:augment}. The "Skin" label is much more common compared to the "Lesion" label. This causes the network to favor optimizing for the "Skin" label over the "Lesion" label which reduces segmentation performance. In order to adjust for this imbalance class weighting is used in the final network layer. Class weights are calculated by dividing class pixels by total pixels averaged over all labels.

\begin{table}
\centering
\begin{tabular}{ll} 
\toprule
\textbf{Augmenter}      & \textbf{Range}  \\ 
\hline
Random X, Y Reflection                                 & True            \\
Random X Translation & [-100 100]px      \\
Random Rotation                                        & [-30 30]$^{\circ}$    \\
Random X, Y Scaling  & [0.75,1.5]~     \\
\bottomrule
\end{tabular}
\caption{Data Augmentation Settings}
\label{tab:augment}
\end{table}

\subsection{Training Data Experiment}
This experiment serves to answer the first subquestion:
\begin{enumerate}
\item How does training data impact specific and general segmentation performance?
\end{enumerate}

Figure \ref{fig:SGN3} shows a plot of the network structure used in this experiment. This SGN3 network is a basic U-net structure with an encoder depth of 3 with 2 convolutional layers per depth level. It is trained from scratch thrice on three different data sets. All other parameters are kept the same. 70\% of the data set is used to train the network, the remaining 30\% is used as test set. After training specific performance metrics are measured on the test set. The ISIC data set is not used for training. Instead it serves as a hold-out set. After training network performance is tested on this set, which serves as an indicator of network generalization performance. 

\begin{figure}[h]
	\begin{center}
		\includegraphics[width=420pt]{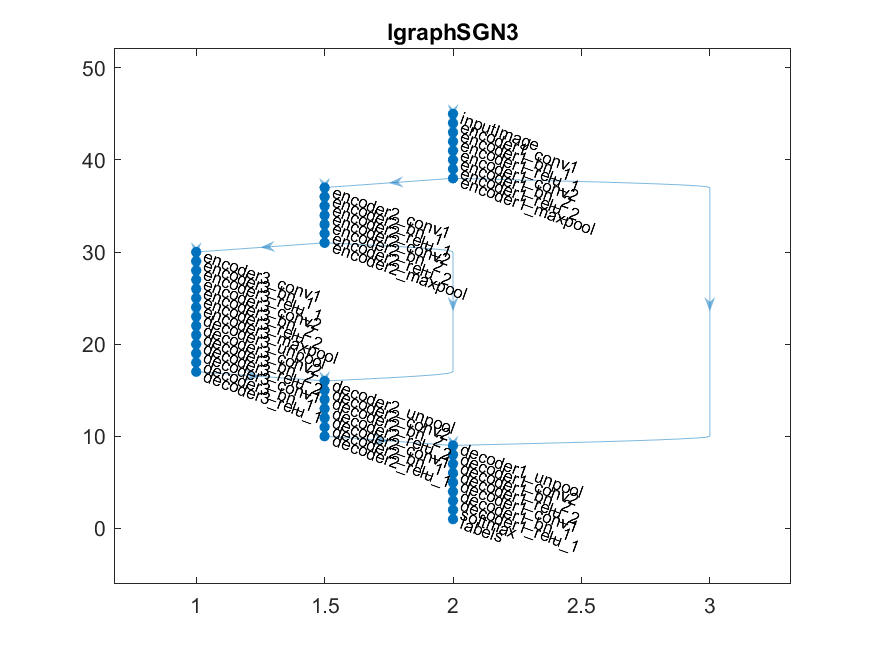}
	\end{center}
	\caption{Plot of SGN3 Network Structure used in the first experiment}
	\label{fig:SGN3}
\end{figure}

Table \ref{tab:trainingdata} contains data set specifications of the three sets used in this experiment. The data in the \textit{cropped} Training Data experiment contains cropped versions sampled from the original UT data set. Each original image in the UT set is sampled 10 times. Each image contains a crop of different, or a combination of, interesting regions. 

This data is further enhanced for the \textit{augmented} Training Data experiment. Histogram equalization, median filtering and edge enhancement pre processing filters are applied to each training images. Five types of noise are applied ten times to a copy of each image for a total of 50 additional images per original image. Table \ref{tab:noisespec} contains specifications on the type and strength of noise. $P(x,y,c)$ is the pixel intensity at width $x$, height $y$ and channel $c$. $|P|$ it the average pixel intensity of the image.

\begin{table}
\centering
\begin{tabular}{llll} 
\toprule
\textbf{Training Data Experiment} & \textbf{Data Sets} & \textbf{Augmentation}      & \textbf{Total Images}  \\ 
\hline
\textit{original}                                                         & UT                 & None                       & 41                     \\
\textit{cropped}                        & UT + DFCN          & Cropping                   & 536                    \\
\textit{augmented}                                                        & UT + DFCN          & Cropping, Filtering, Noise & 27336                  \\
\bottomrule
\end{tabular}
\caption{Specifications of training data used in the first experiment}
\label{tab:trainingdata}
\end{table}


\begin{table}
\centering
\begin{tabular}{lcc} 
\toprule
\textbf{Noise Type}          & \textbf{Times applied} & \textbf{Strength}  \\ 
\hline
Gaussian                                                    & 10                     & $\sigma^2=0.01, \mu = 0$ \\
Gaussian - Local Variance & 10                     & $\sigma^2=0.01 \cdot \frac{P(h,w,c)}{|P|}, \mu = 0$                    \\
Poisson                                                     & 10                     & \href{https://nl.mathworks.com/help/images/ref/imnoise.html#mw_226e1fb2-f53a-4e49-9bb1-6b167fc2eac1}{Poisson algorithm}                   \\
Speckle                   & 10                     & Multiplicative with $\sigma^2=0.04, \mu = 0$                   \\
Salt \& pepper                                              & 10                     & Noise density $\rho= 0.05$                   \\
\bottomrule
\end{tabular}
\caption{Specifications of Noise Types in the \textit{augmented} Training Data experiment}
\label{tab:noisespec}
\end{table}

Training options for this experiment are listed in table \ref{tab:opttrain}. Stochastic gradient descend with momentum 0.9 is used as the optimization algorithm. A mini-batch size of 1 is very small but necessary to run the script on a lower end computer. Every epoch all images in the training set are shuffled to prevent adaptations of the network to the order in which images are presented. All scripts are going to be shared on the github repository of the first author.

\begin{table}
\centering
\begin{tabular}{ll} 
\toprule
\textbf{Training option} & \textbf{Setting}  \\ 
\hline
Algorithm                                               & SGDM              \\
Momentum              & 0.9               \\
Learning rate                                           & 0.003             \\
L2 Regularization     & 0.0005            \\
Maximum Epoch                                           & 100               \\
Mini-batch Size       & 1                 \\
Validation Data                                         & Test Set          \\
Validation Patience   & 10 Epochs         \\
Shuffling~                                              & Every Epoch       \\
\bottomrule
\end{tabular}
\caption{Training options used in the Training Data experiment}
\label{tab:opttrain}
\end{table}

\subsection{Network structure Experiment}
This experiment attempts to answer the second and third subquestions:
\begin{enumerate}
\item How does network structure impact specific and general segmentation performance
\item How does training method impact specific and general segmentation performance?
\end{enumerate}

\begin{figure}[h]
        \centering
        \begin{subfigure}[b]{200pt}
            \includegraphics[width=200pt]{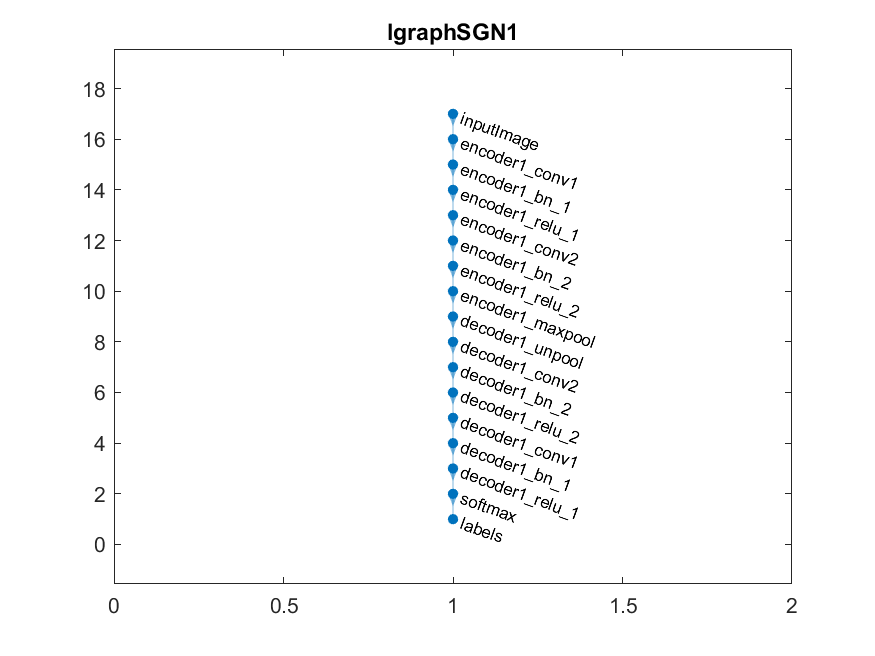}
            \caption{SGN1 Network structure}
            \label{fig:SGN1}
        \end{subfigure}
    \begin{subfigure}[b]{200pt}
        \includegraphics[width=200pt]{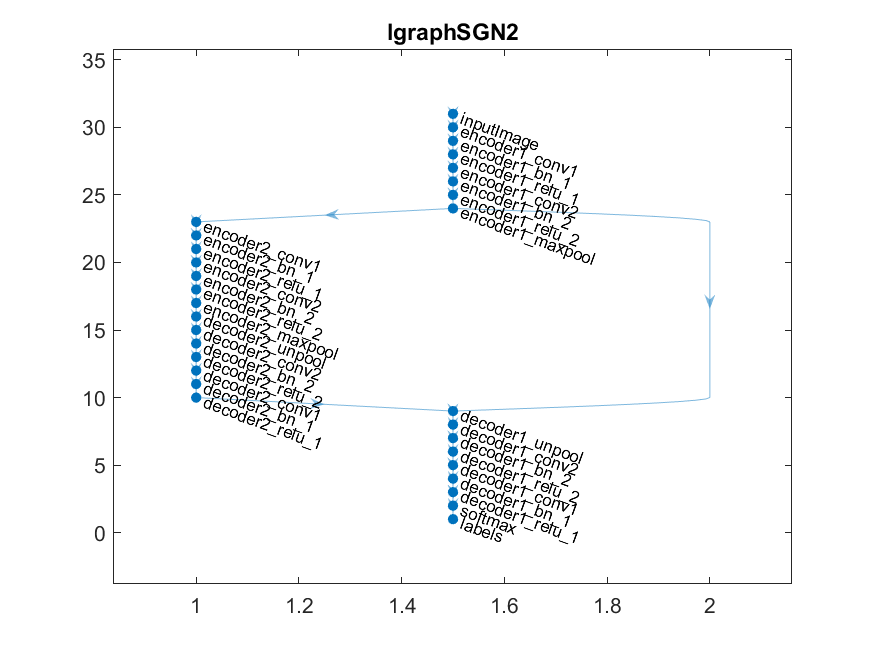}
        \caption{SGN2 Network structure}
        \label{fig:SGN2}
    \end{subfigure}
    \begin{subfigure}[b]{200pt}
        \includegraphics[width=200pt]{images/lgraphSGN3_Structure.png}
        \caption{SGN3 Network structure}
        \label{fig:SGN3}
    \end{subfigure}
    \begin{subfigure}[b]{200pt}
        \includegraphics[width=200pt]{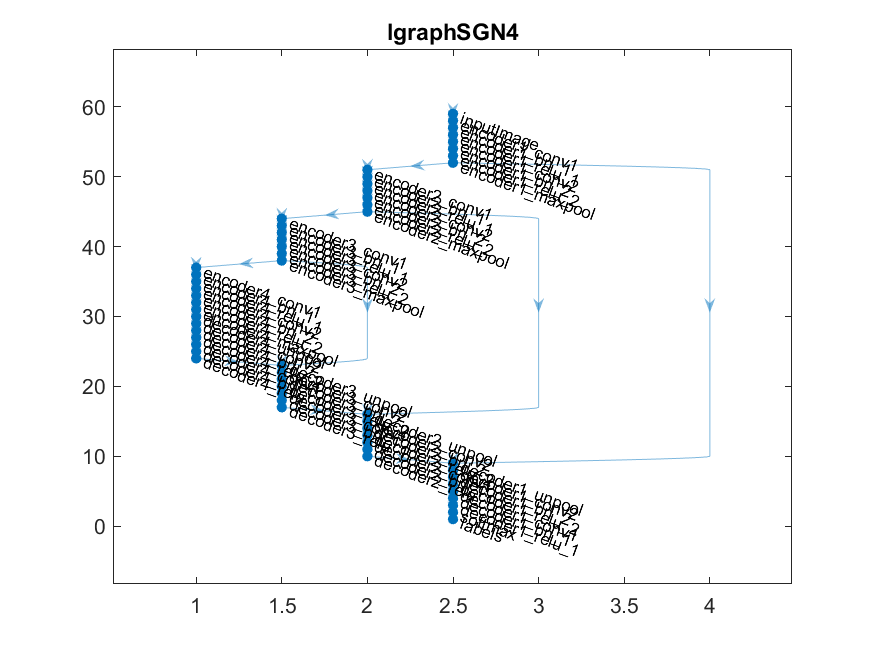}
        \caption{SGN4 Network structure}
        \label{fig:SGN4}
    \end{subfigure}
    \begin{subfigure}[b]{200pt}
        \includegraphics[width=200pt]{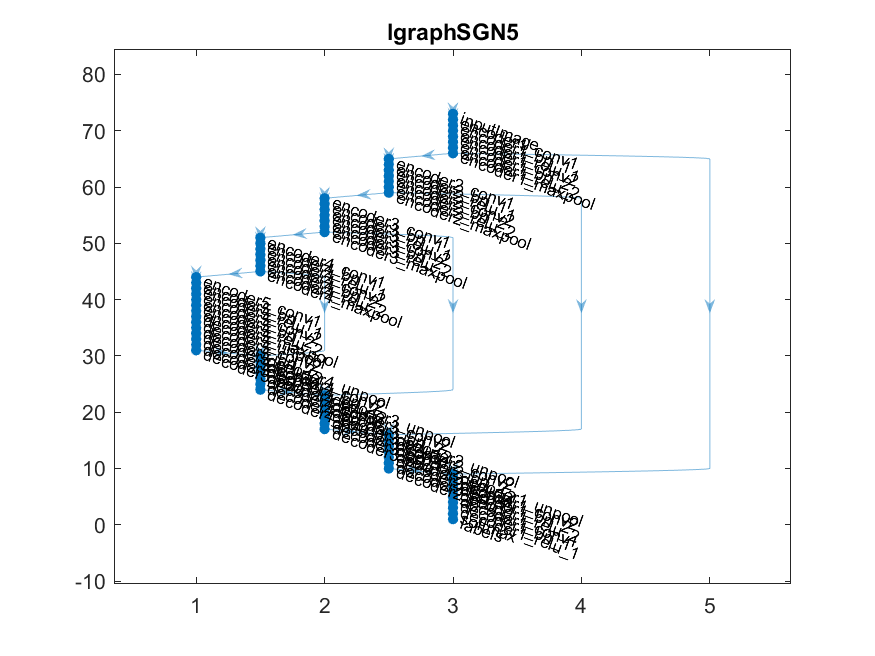}
        \caption{SGN5 Network structure}
        \label{fig:SGN5}
    \end{subfigure}
    \begin{subfigure}[b]{200pt}
        \includegraphics[width=200pt]{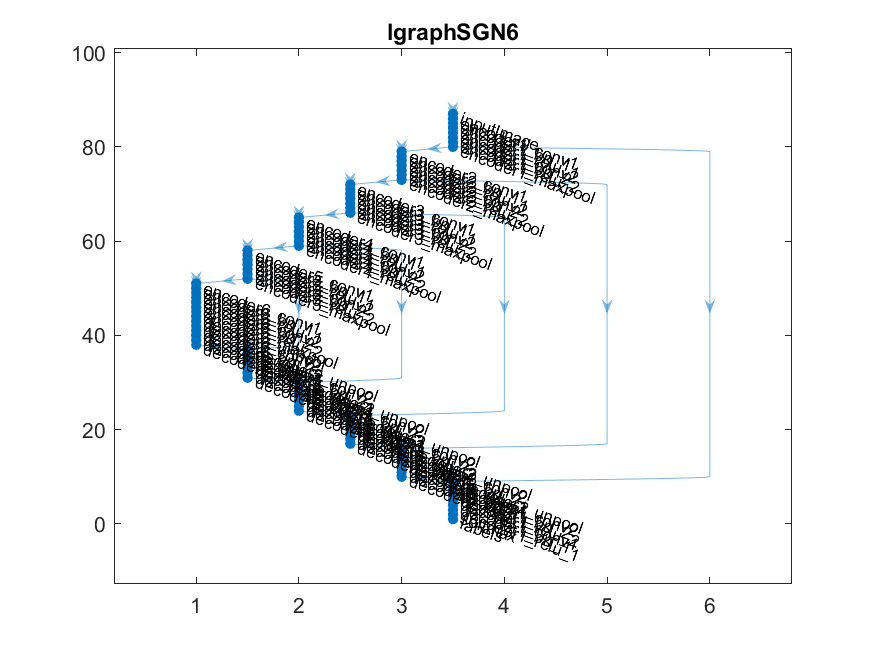}
        \caption{SGN6 Network structure}
        \label{fig:SGN6}
    \end{subfigure}
    \caption{Plots of SGN1-6 Networks used in the Network Structure experiment}
    \label{fig:SGNplot}
\end{figure}

\begin{figure}[h!t]
    \centering
    \begin{subfigure}[b]{200pt}
        \includegraphics[width=200pt]{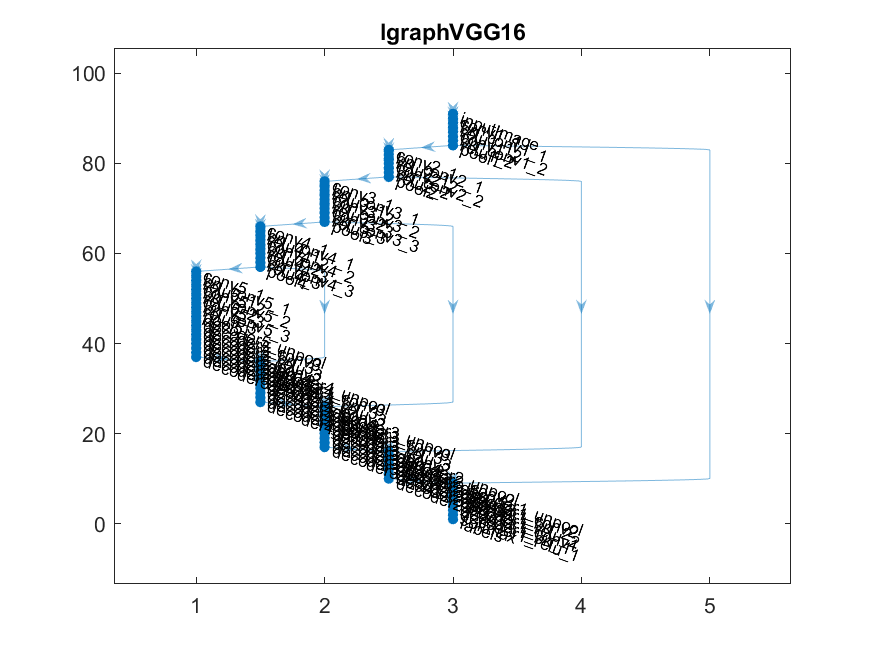}
        \caption{VGG16 Network structure}
        \label{fig:VGG16}
    \end{subfigure}
        \begin{subfigure}[b]{200pt}
        \includegraphics[width=200pt]{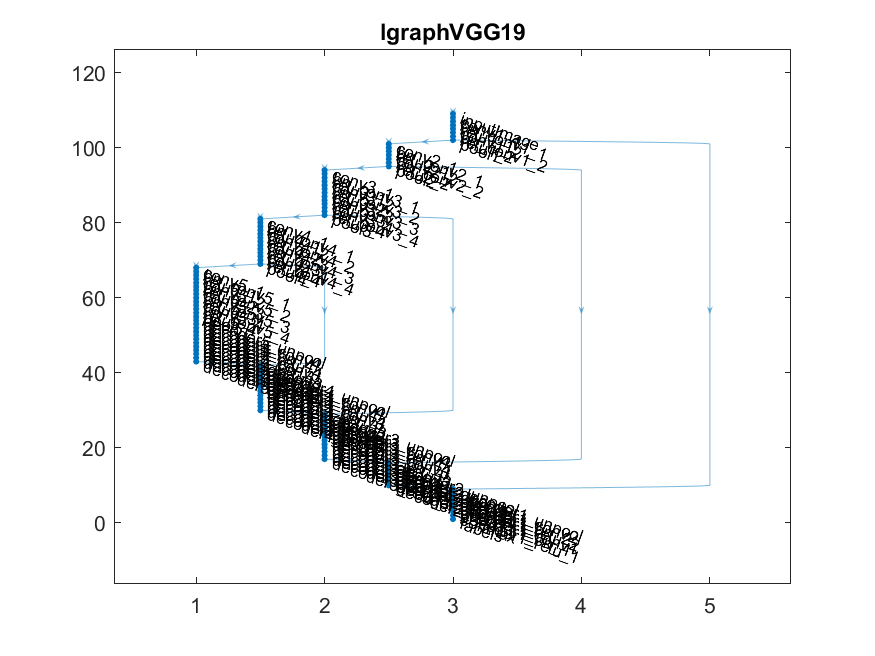}
        \caption{VGG19 Network structure}
        \label{fig:VGG19}
    \end{subfigure}
    \begin{subfigure}[b]{200pt}
        \includegraphics[width=200pt]{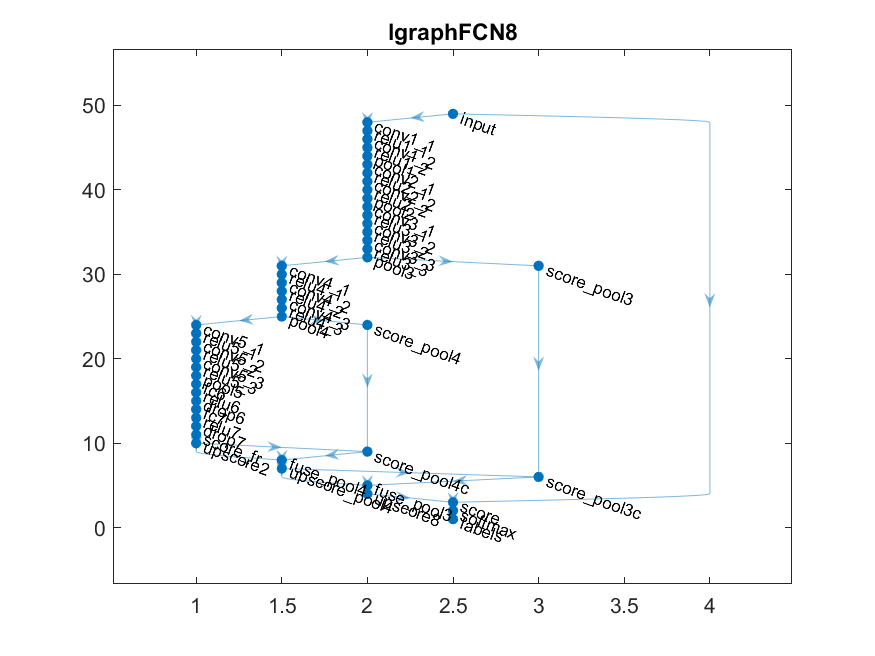}
        \caption{FCN8 Network structure}
        \label{fig:FCN8}
    \end{subfigure}
    \begin{subfigure}[b]{200pt}
        \includegraphics[width=200pt]{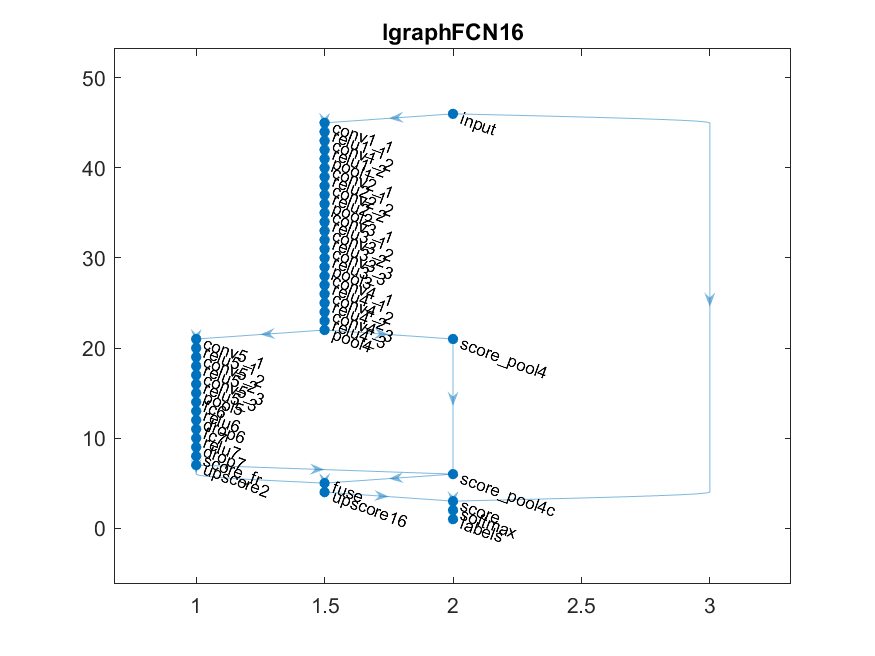}
        \caption{FCN16 Network structure}
        \label{fig:FCN16}
    \end{subfigure}
    \begin{subfigure}[b]{200pt}
        \includegraphics[width=200pt]{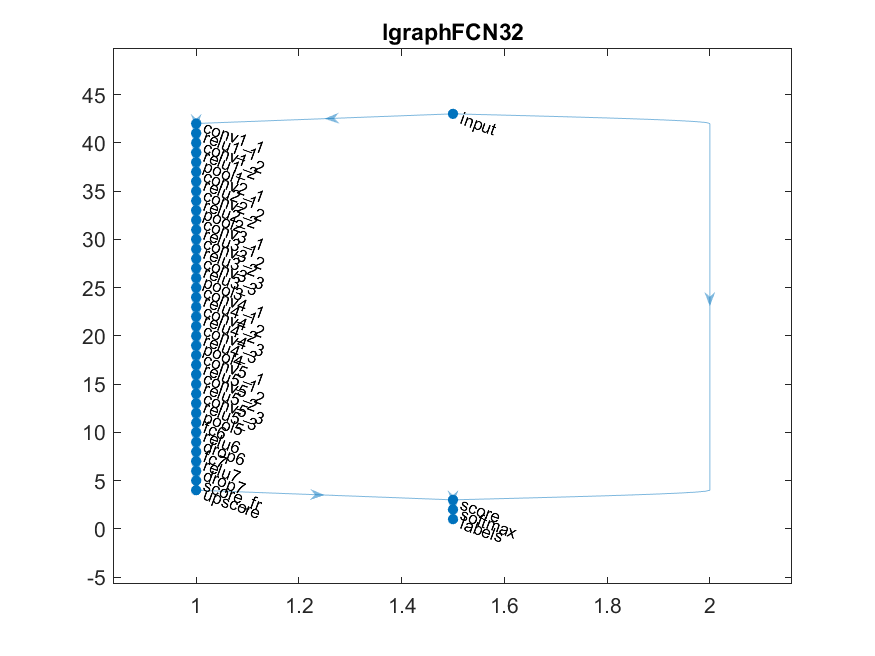}
        \caption{FCN32 Network structure}
        \label{fig:FCN32}
    \end{subfigure}
    \begin{subfigure}[b]{200pt}
        \includegraphics[width=200pt]{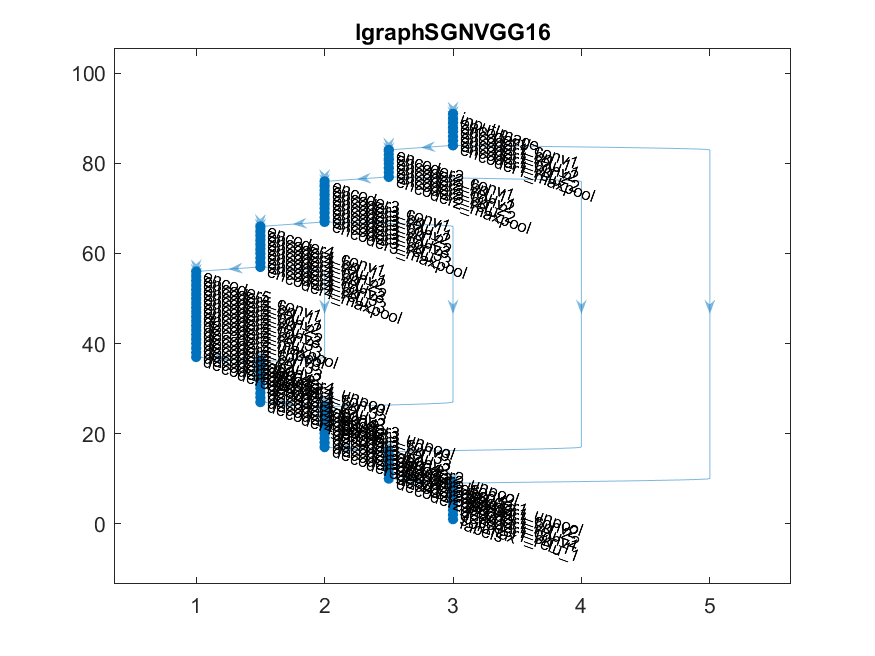}
        \caption{SGNVGG16 Network structure}
        \label{fig:SGNVGG16}
    \end{subfigure}
    \caption{Plots of network structures used in the Network Structure experiment}\label{fig:ActivationFunc}
\end{figure}

In this experiment multiple different network structures are trained. Only the structure differs, all other parameters are kept constant. Almost all training options are the same as those used in the Training Data experiment. To reduce computation time the number of maximum epochs have been reduced to 50. To improve network performance results the early stopping condition has been relaxed to 25 validation iterations. All networks are based on the U-net encoder decoder structure. Figures \ref{fig:SGN1} to \ref{fig:SGNVGG16} show plots of the networks trained in this experiment. 

Networks SGN1-6 are straightforward U-net implementations. SGN1 has an encoder depth of 1, SGN6 has a depth of 6. Each encoder section contains two convolution layers with batch normalization, ReLu activation layers and a max pooling layer. Each decoder section receives information from the decoding section below and its corresponding encoding section. The weights of all SGN networks are randomly initialized. 

VGG16 and VGG19 are similar networks based on the 16 layer deep VGG16 classification network \cite{Simonyan2015VERYRECOGNITION}. VGG16 has an encoder depth of 5. Its first two encoding sections have two convolution layers, the last three sections have three convolution layers. VGG19 is based on the pre trained VGG19 network. It is slightly deeper than the VGG16 network. The weights of both VGG16 and VGG19 are initialized using weights of their original counterparts, their decoder weights are initialized randomly.

Introduced by J. Long et al. \cite{LongFullySegmentation}, the FCN network structures are similar to the SGN networks. Each decoder section of the FCN structures take information of the decoder section below and a pooled combination of the decoder section below and the input of the encoding section on the same depth. The encoder depth of FCN32, FC16 and FC8 is 1, 2 and 3 respectively. Compared to the SGN networks the skip connections in the FCN networks pass information earlier. The Weights of the FCN8, 16 and 32 networks are initialized using VGG16 weights. 

SGNVGG16 has exactly the same structure as VGG16. It has the same encoder depth of 5 with the first two layers containing two convolution layers per section and the other sections containing three. The only difference are the weights which are all randomly initialized. 

All networks are trained on the same data set used in the \textit{cropped} Training Data experiment, see table \ref{tab:trainingdata}. This data set contains a total of 536 images. 

\subsection{Performance Metrics}
Performance of the networks in both experiments is tested on the test and hold-out set. The training data set in the Training Data experiment varies between each of the tested networks. Since the test set is 30\% of total training data this means the test set also varies. Networks in the Network Structure experiment all use the same training data and test sets. The hold-out set consists of 10 images from the ISIC database and remains constant throughout all experiments.

Performance on the test set gauges the networks' specific performance. Performance on the hold-out set shows the networks capability to generalize to new data, or generalization performance. Performance metrics used are the confusion matrix and the label specific accuracy, intersection over union (IoU) and Mean Boundary F1 Score which is the F1 score calculated at label boundaries. This provides information of a networks segmentation performance at lesion boundary regions. This is very useful when aiming to apply post processing to the networks output. Training time is also taken into consideration.

\section{Results and Discussion}

\label{ch:discussion}
\subsection{How does training data impact segmentation performance?}
Due to computational constraints the performance of the SGN3 network on the Augmented set can not be tested. The SNG3 network trained on the \textit{original} data is referred to as the \textit{original} network. The SNG3 network trained on the \textit{cropped} data set is referred to as the \textit{cropped} network. Several observations are made. Compared to the \textit{original} network the skin label accuracy on the test set of the \textit{cropped} network is 6.65\% better, but the lesion accuracy is 14.42\% worse. On the contrary the skin label IoU of both networks are similar but the lesion IoU of the \textit{cropped} network is 82.49\% better. The \textit{original} network has better Mean BF1 scores on both the skin and lesion label with 11.53\% and 18.02\% respectively. 

\begin{figure}[h!t]
    \centering
    \begin{subfigure}[b]{200pt}
        \includegraphics[width=180pt, trim={1cm 1cm 1cm 1cm},clip]{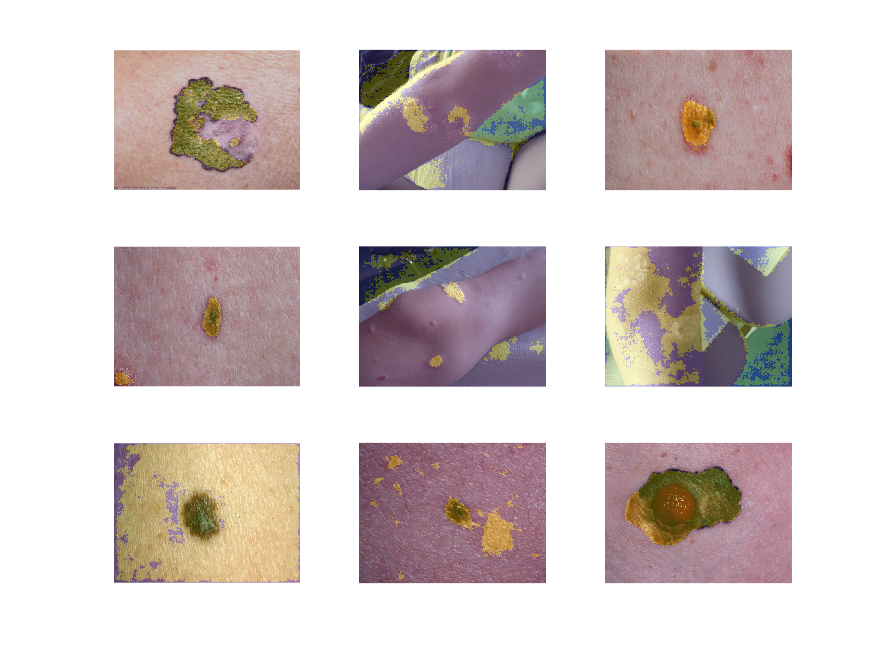}
        \caption{\textit{original} network}
        \label{fig:ogex}
    \end{subfigure}
    \begin{subfigure}[b]{200pt}
        \includegraphics[width=180pt, trim={1.4cm 1.4cm 1.4cm 1.4cm},clip]{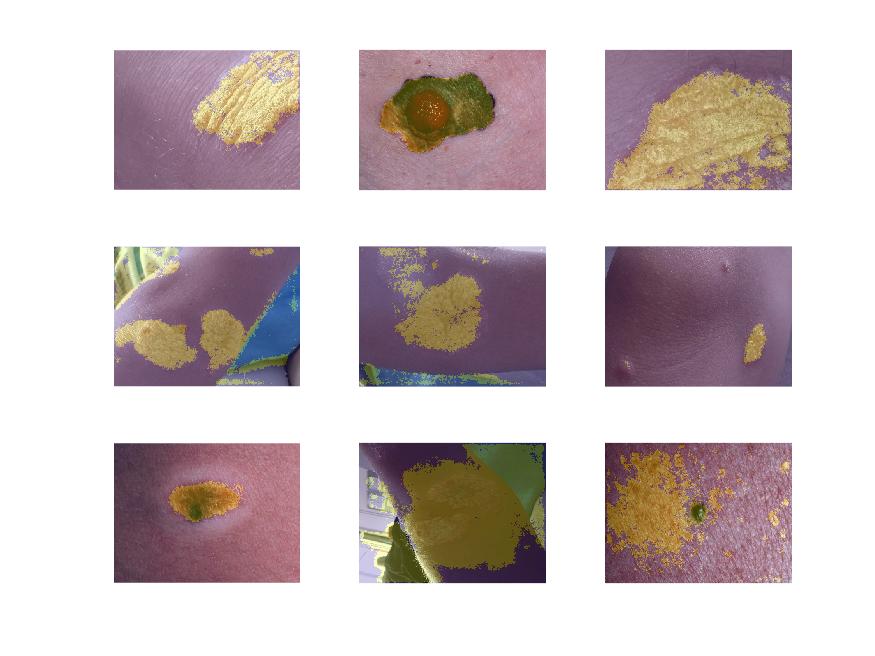}
        \caption{\textit{cropped} network}
        \label{fig:cropex}
    \end{subfigure}
    \caption{Segmentation examples of \textit{original} and \textit{cropped} networks}\label{fig:ogcropex}
\end{figure}

On the ISIC set these differences are less pronounced. On this set the \textit{cropped} network has 3.61\% and 2.21\% better skin and lesion accuracy. The skin and lesion IoU of the \textit{cropped} network is 3.86\% and 8.13\% better. The \textit{original} network has better Mean BF1 performance with 2.14\% and 3.18\% increase over the \textit{cropped} network for skin and lesions. Figure \ref{fig:ogex} and \ref{fig:cropex} show a comparison of network performance and segmentation examples of the test set. 

\begin{figure}[h]
	\begin{center}
		\includegraphics[width=420pt]{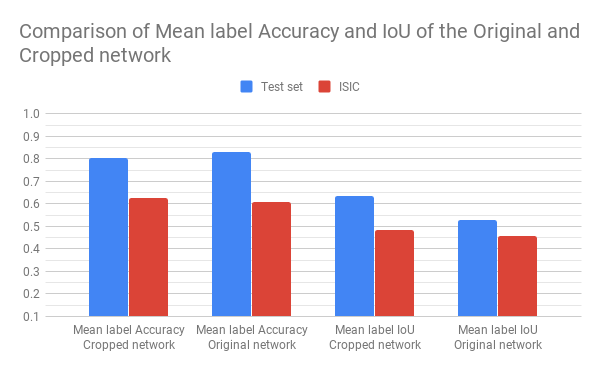}
	\end{center}
	\caption{Mean label Accuracy and IoU of \textit{original} and \textit{cropped} network}
	\label{datacomp}
\end{figure}

Figure \ref{datacomp} shows a comparison of the Accuracy and IoU averaged over both labels of the \textit{original} and \textit{cropped} network. The \textit{cropped} network outperforms the \textit{original} network in every IoU metric. Accuracy of the \textit{cropped} network is 0.028 worse on the test but 0.019 better on the ISIC set. The \textit{cropped} network generalizes better to new data.

A skin lesion segmentation model needs high precision to function in a clinical setting because of the serious consequences and cost of false positives. With this in mind the \textit{cropped} network arguably outperforms the \textit{original} network, despite slightly lower test set accuracy. The \textit{original} network correctly assigns more pixels but the \textit{cropped} network more often assigns groups of pixels correctly. The \textit{cropped} network also appears more robust to non-uniform skin textures and variations in image lighting. These observations make the \textit{cropped} network more suitable to identify lesion area. 

The Mean BF1 scores of the networks do not fully support this conclusion. The \textit{original} network scores better on the test set and ISIC set for both labels. However this difference is small. Averaged over both labels the difference in mean BF1 score is 0.0069 and 0.0072 on the test and ISIC set respectively. 

It is important to emphasize the difference in size of the test sets. The test sets consist of 30\% of total training data. The test set for the \textit{original} network contains 12 images, the test set for the \textit{cropped} network contains 161. Because of this conclusions drawn from these observations have arguable relevance.

From the experimental results can be concluded that: (1) More training data is related to better segmentation performance. (2) More training data is related to better network generalization. 


\begin{figure}[h!t]
	\begin{center}
		\includegraphics[width=420pt]{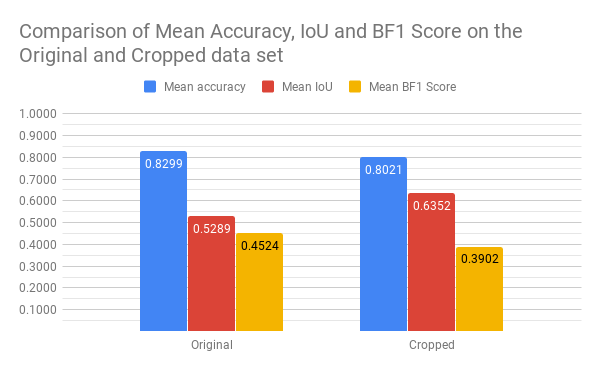}
	\end{center}
	\caption{Test set accuracy and segmentation examples on the \textit{cropped} data set}
	\label{fig:cropacc}
\end{figure}

\subsection{How does network structure impact segmentation performance?}
When comparing all performance metrics, the FCN32 and FCN16 perform best. FCN32 has a mean label accuracy, IoU and BF1 score of 0.962, 0.810 and 0.551 respectively. FCN16 has 0.961, 0.810 and 0.563 for the same metrics, both on the test set. FCN32 slightly outperforms FCN16 in the accuracy metric. FCN16 has a slightly higher BF1 Score. On the ISIC set FCN32 achieves an mean label accuracy, IoU and BF1 score of 0.759, 0.658 and 0.548 respectively. For each of these metrics FC16 achieves 0.820, 0.701 and 0.509. FCN16 outperforms FCN32 with 8.08\% and 6.49\% in mean label accuracy and IoU. FCN32 has a 7.68\% increased BF1 score compared to FCN16. FCN16 appears to segment more coarsely compared to FCN32. This is likely a result of the difference in network structure. FC16 has an encoder depth of 2, FCN32 has a depth of 1. This supports the intuitive belief that networks with larger encoder depths look for finer grained structures. This is likely the reason FCN32 has a better mean BF1 score but worse accuracy and IoU performance. BF1 score is more heavily influenced by falsely labeled pixels in the label boundary region. Compared to FCN32, FCN16 has coarser segmentation behavior which would result in more erroneously labeled boundary pixels. 

Surprisingly FCN8 achieves very poor scores on almost all performance metrics, scoring much lower than the  FCN16 and FCN32 networks. FCN8 has a similar architecture to FCN16 and 32 but has one added encoding section. All networks have the same amount of convolution layers. Compared to other metrics, FCN8 scores remarkably well on lesion label accuracy. On the Test set FCN8 scores 0.707 and 0.746 on the skin and lesion accuracy metric. Interestingly the lesion accuracy is higher than the skin accuracy. Only the very shallow SGN1 network has the same result. On the ISIC set these scores decrease to 0.308 and 0.861. The skin label accuracy decreases dramatically with 0.399. However lesion accuracy increases with 0.114. This shows that the significant decrease of performance of the FCN8 network might be explained by class weighting. For the FCN8 structure the magnitude class weighting could have been to high causing the network to heavily favor optimizing for lesion accuracy, decreasing its overall performance.

On average SGN structures perform much worse compared to FCN structures across all metrics, especially when comparing network generalization performance on the ISIC set. This decrease in performance could be explained by the difference in insertion position of the skip architectures. SGN structures use skip architectures that combine information from the decoder section below with the output of the encoder section on the same depth. Skip architectures in FCN structures also use information from the decoder section below but combine it with information from a pooling layer. This pooling layer takes information from the start of the encoder section on the same depth and combines it information of the decoder section below. This combination of less processed information pooled with more heavily processed information are likely the reason FCN structures perform better.

Training time also heavily differs between the SGN and FCN structures. FCN8, 16 and 32 have trained for 4:50, 3:46 and 3:49 respectively. Disregarding the shallow SGN1 and SGN2 networks, FCN networks train on average 48:20 longer compared to SGN networks. Training of the FCN networks also behaved much more erratic, with frequent changes in loss and accuracy during training orders of magnitude bigger compared to the SGN networks. This is likely the reason for the increased average training time since all networks stopped training at some point by achieving the early stopping condition. Training of the FCN networks was more erratic and therefore took longer to achieve the stopping condition. This property of the FCN networks can also indicate a higher variation when revising the experiment.

The VGG structures also perform well. VGG16 has a mean label accuracy, IoU and BF1 Score of 0.936, 0.775 and 0.539. VGG19 scores 0.850, 0.756 and 0.520 in these categories. Interestingly the VGG16 network outperforms its deeper VGG19 counterpart. When looking at individual label accuracy VGG16 scores 0.957 and 0.914 for the skin and lesion label respectively. VGG19 achieves the highest skin label accuracy of 0.993 but a more modest 0.701 for lesion accuracy. This effect is more pronounced on the ISIC set. Label specific accuracy for VGG16 are 0.906 and 0.644 for skin and lesion respectively. VGG19 scores a 0.948 and 0.326 skin and lesion accuracy. This is likely due to the difference in size of the skin and lesion labels. For larger networks the class weighting has not dissuaded the network enough to fit for both classes. Overfitting could also have occured but is less likely due to the excellent skin accuracy of both networks on the ISIC set. These networks likely require either more regularization during training to achieve maximum performance.

\begin{figure}[h]
	\begin{center}
		\includegraphics[width=420pt]{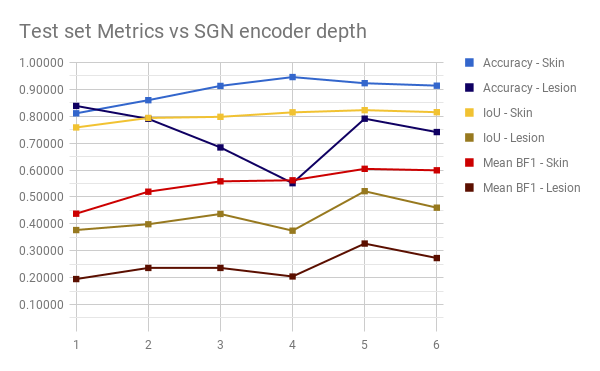}
	\end{center}
	\caption{Test set performance of SGN1-6}
	\label{fig:encode}
\end{figure}

\begin{figure}[h!t]
	\begin{center}
		\includegraphics[width=420pt]{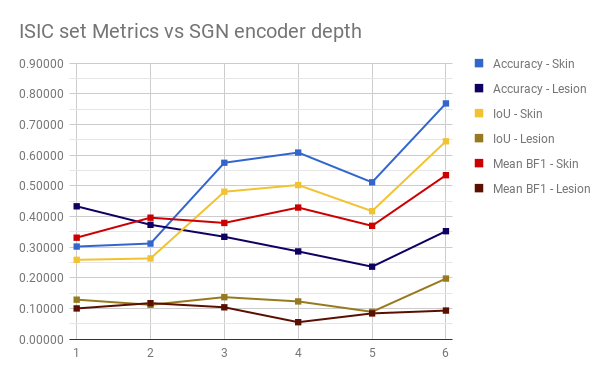}
	\end{center}
	\caption{ISIC set performance of SGN1-6}
	\label{fig:encodeISIC}
\end{figure}

Another interesting observation is the effect of encoder depth on the performance of the SGN networks, shown in figure \ref{fig:encode}. IoU, Mean BF1 score and skin accuracy on the test set all have upward trends when increasing encoder depth, with maximum performance arguably achieved by the SGN5 network. Interestingly the lesion accuracy has a downward trend with SGN4 performing much worse than SGN1 to 3. Comparing these observations to the performance of the SGN networks on the ISIC set in figure \ref{fig:encodeISIC} shows encoder depth is positively related to network performance. Performance of SGN on the ISIC set is much worse compared to SGN4 and 6. This shows SGN5 is slightly overfitting on the test set. The same downward trend in lesion accuracy exists on the ISIC set. This indicates that the training algorithm judges the best weight changes each iteration to be those that reflect a more positive change in skin accuracy in more shallow networks. A deeper network is better able to optimize for lesion accuracy as well.


\begin{figure}[h!t]
	\begin{center}
		\includegraphics[width=420pt]{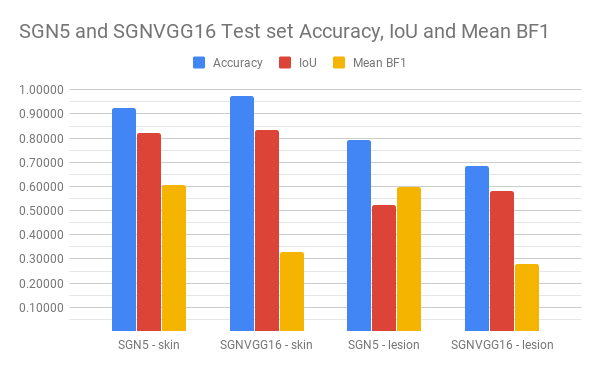}
	\end{center}
	\caption{Comparison of performance of SGN5 and SGNVGG16 on the test set}
	\label{fig:comp1}
\end{figure}

\begin{figure}[h!t]
	\begin{center}
		\includegraphics[width=420pt]{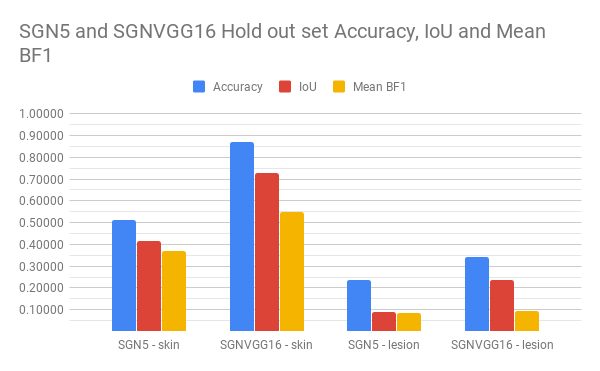}
	\end{center}
	\caption{Comparison of performance of SGN5 and SGNVGG16 on the ISIC set}
	\label{fig:comp2}
\end{figure}

Figure \ref{fig:comp1} and \ref{fig:comp2} show a comparison between the performance of SGN5 and SGNVGG16 on the test and ISIC set. On the test set the SGN5 network performs better compared to the SGNVGG16 network. However the SGNVGG16 outperforms the SGN5 network when tested on the ISIC set. This indicates the SGN5 network is overfitting on the training data and has worse generalization performance than SGNVGG16. The SGN5 and SGNVGG16 networks are very similar. Both have an encoder depth of five. SGN5 has two convolutional layers per encoder section. SGNVGG16 is modeled after the pre trained VGG16 network and has two convolutional layers in the first two layers and three convolutional layers in the last three layers. To explain whether the difference in performance of these networks is due to this difference in structure or to the random nature of network training requires more investigation. 


From these observations the following conclusions can be drawn:
(3) Increasing the number of convolution layers per encoding section is more effective than increasing network encoder depth. (4) When using skip architectures, combining information from a deep layer with relatively unprocessed information from a higher layer is more effective than combining it with more heavily processed information. (5) Training FCN network structures using early stopping regularization increases training time. (6) For SGN structures increased encoding depth is related to increased generalization performance.

\subsection{How does training method impact segmentation performance?}
Training a network that uses weights of a pre trained network is called transfer learning. Training a network that uses randomly initialized weights is referred to as training from scratch. 

\begin{figure}[h!t]
	\begin{center}
		\includegraphics[width=420pt]{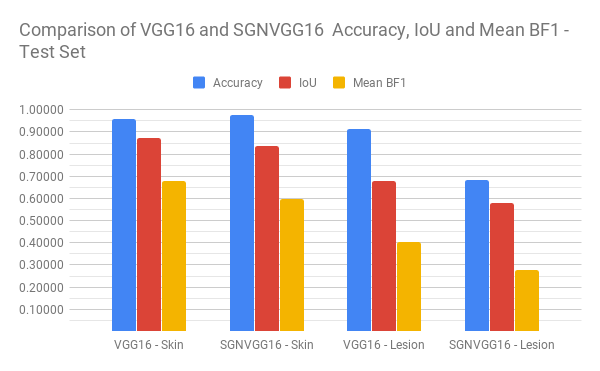}
	\end{center}
	\caption{Test set Performance metrics Comparison of VGG16 and SGNVGG16 networks}
	\label{fig:methcomptest}
\end{figure}

\begin{figure}[h!t]
	\begin{center}
		\includegraphics[width=420pt]{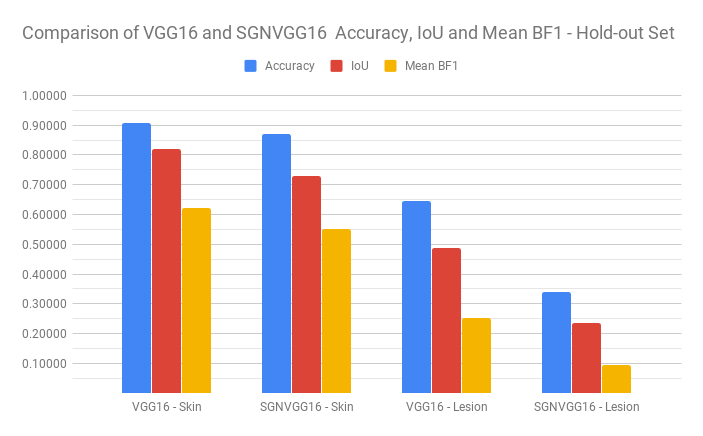}
	\end{center}
	\caption{ISIC set Performance metrics Comparison of VGG16 and SGNVGG16 networks}
	\label{fig:methcomphold}
\end{figure}

Figure \ref{fig:methcomptest} and \ref{fig:methcomphold} show a comparison of test and ISIC set performance metrics of the VGG16 and SGNVGG16 networks. The only difference between these networks is the initialization method of their weights, their structures are identical. SGNVGG16 weights are initialized randomly, VGG16 weights are initialized using weights of the well-known VGG16 network. Since the original VGG16 network is a classification network only the encoder weights of the version used in this experiment are transferred. VGG16 decoder weights are initialized randomly.

\begin{figure}[h!t]
	\begin{center}
		\includegraphics[width=420pt]{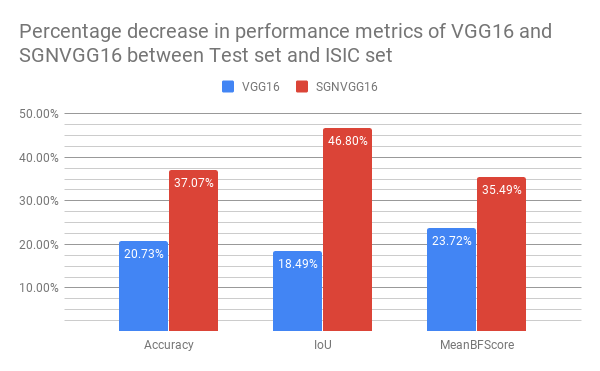}
	\end{center}
	\caption{Comparison of the percentage decrease in label-averaged performance metrics between the test and ISIC set}
	\label{fig:traindecr}
\end{figure}

On the test set, the VGG16 network outperforms the SGNVG16 network in every performance metric except for skin label accuracy. Compared to VGG16 the skin label accuracy of the SGNVGG16 is 0.068 better. When testing on the ISIC set VGG16 performs much better by every metric. The biggest difference is the performance of the networks when labeling lesion pixels. When comparing the ISIC set metrics to the test set metrics VGG16 has a 20.73\%, 18.49\% and 23.72\% reduction in mean accuracy, IoU and BF1 Score respectively. SGNVGG16 has a 37.07\%, 46.80\% and 35.49\% reduction in mean accuracy, IoU and BF1 Score. Figure \ref{fig:traindecr} shows this difference. From this can be concluded the VGG16 is better able to generalize to new data. This supports the notion in the literature that suggests transfer learning increases network generalization\cite{YosinskiHowNetworks}. 

Another observation is the difference in training time when comparing VGG16 to SGNVGG16. VGG16 took 2:54:29 to train, SGNVGG16 took 3:44:47 using the same training options. This is a significant reduction of 50 minutes and 18 seconds.

The FCN networks also have weights based on the pre trained VGG16 network. As with the VGG16 network used in this experiment only the encoder weights are transferred and decoder weights are randomly initialized. The VGG19 network also uses encoder weights transferred from its pre trained counterpart. All FCN and VGG networks have above average performance metrics. The extent of the influence transfer learning has on performance is difficult to identify. However it is safe to say transfer learning has aided the performance of these networks. 

From these observations can be concluded: (7) when comparing transfer learning to learning from scratch,  transfer learning yields better specific and general performance even when weights are only partially transferred. (8) The increase of performance by utilizing transfer learning is greater on general performance than specific performance. (9) Utilizing transfer learning significantly reduces training time. 


\subsection{Shortcomings and Clinical Implementation}
Training neural networks is a stochastic process. Training the same network keeping all hyperparameters constant yields different results every time. In order to validate the conclusions drawn from this experiment insight into the variance of this experiment is necessary. Especially the FCN structures have very erratic loss functions while training, possibly indicating a high variance. Valuable future research would be investigating the variance of the experiments by repeating them multiple times. This is necessary to validate the conclusions drawn in this paper.

Other future research could include experimenting with different training options. Experimental results provided in this paper are obtained using very similar training options. Exploring the impact of training options on network structure is necessary to validate the best performing network structures. SGN structures could possibly outperform FCN structures if given more training time and a more relaxed early stopping condition. More research should be conducted into the underlying reasons of the excellent performance of the FCN networks. This could be further comparison between SGN, FCN and similar network structures. Or research into the significant difference in training behavior of the SGN and FCN structures.  

Testing networks on the Augmented sets could unfortunately not be done due to computational constraints. Experimenting with the Augmented set and further research into the impact of training data is necessary to validate the conclusions drawn from the Training Data experiment. Further research could include comparing the performance of different structures trained on a large number of different data sets. To more accurately discern the consequence on network performance of different pre processing filters, a similar structure should be trained on differently pre processed data sets. Researching the consequences addition of noise yields on network performance can also yield valuable insight. 

In order for this system to be implemented in a clinical environment it needs additional work. The accuracy of the segmentation achieved in this paper is not enough to pass clinical performance tests and should be improved. This improvement can come from changes in the machine learning structure or from added post processing. This system could be employed to do a rough segmentation of image material and use other segmentation methods like active contouring to achieve a high final segmentation performance. 

In order to accurately track a patients lesion over time a method needs to be devised to present the system with an image taken at the same distance from the skin. If the system does not recognize scale the resulting tracking could be very inaccurate due to changes in distance between measurement images. This could be done using any distance measuring tool, for instance a laser that measures distance attached to the camera. Another method to solve this would be to implement a subsystem that relates skin lesion size to other visual clues like freckles, tattoos or other irregularities. 

Other functionality that is valuable to add for use in a clinical setting is a severity classifier. This classifier would give insight into the severity of a lesion or tumor by means of shape and texture analysis. Generally non-uniform shapes and textures are related with more active and severe tumors. To further increase accuracy of the severity classification thickness and relative redness analysis functionality can be added.  Lesion thickness is difficult to measure and therefore difficult to monitor, making it a useful addition to a clinical lesion model. In order to add this functionality some sort of 3D data should be gathered. 

As the first experiment suggests, maximum performance of the networks in this paper can only be achieved when a large set of training is available. If qualitatively labeled data remains sparse. research into different machine learning models should be conducted. Less complex methods like support vector machines can yield better performance on smaller data sets. These methods also reduce the computational cost of network training and inference time. A decrease in training time can aid in development and research. A lower inference time is a valuable property in clinical settings, where real time processing of images is necessary. 

When the lack of labeled data turns out to be the most limiting factor of further improvement, research into unsupervised learning methods can be invaluable. Unsupervised learning methods do not require labeled data. Auto encoders are an example of these type of networks. These networks deconstruct training data into a smaller dimensional representation often called code. It then attempts to reconstruct the original data from the code layer and used back propagation methods to improve this process. Because these networks do not require training data to be labeled finding appropriate training data is much simpler. Another potential method of dealing with sparse training data is the usage of Generative Adversarial Networks to generate new training data. Using GANs for training data generation should be researched further.

\section{Conclusion}
\label{ch-conclusion}
To perform semantic segmentation on skin lesion pictures multiple U-net structures are constructed and tested. The FCN32 and FCN16 structures perform best by label accuracy, IoU and Mean BF1 metrics. All networks using weights partially initialized by the VGG16 network perform above average, except for the FCN8 network. The first experiment yields some observations regarding the first research sub question. The SGN3 network trained on a cropped set generalizes better to new data. The impact of training data on specific and general segmentation performance is: (1) More training data is related to better segmentation performance. (2) More training data is related to better network generalization. 

The impact of network structure on specific and general segmentation performance is researched by means of the second experiment. Observations made are: (3) Increasing the number of convolution layers per encoding section is more effective than increasing network encoder depth. (4) When using skip architectures, combining information from a deep layer with relatively unprocessed information from a higher layer is more effective than combining it with more heavily processed information. (5) Training FCN network structures using early stopping regularization increases training time. (6) For SGN structures increased encoding depth is related to increased generalization performance.

The second experiment also yields observations regarding the research sub question 'How does training method impact specific and general segmentation performance?' From these observations can be concluded: (7) when comparing transfer learning to learning from scratch, transfer learning yields better specific and general performance even when weights are only partially transferred. (8) The increase of performance by utilizing transfer learning is greater on general performance than specific performance. (9) Utilizing transfer learning significantly reduces training time. 

Combining these conclusions yields insight into the research question "\textit{How to segment skin lesion images using a neural network with low available data?}" Of the methods tested in this paper the best approach is using a FCN32/16 network structure with weights transferred from the pre-trained VGG16 network, trained on a representative data set. This set should be pre processed using cropping to increase the information the network can draw from it during training, especially when the set is small.

\bibliographystyle{unsrt}  
\bibliography{template}  

\end{document}